%% file: Cheng_CST.tex
\documentclass[12pt]{article}
\usepackage[margin=1in]{geometry}
\usepackage{cite}
\usepackage{amsmath,amssymb,amsfonts}
\usepackage{algorithmic}
\usepackage{graphicx}
\usepackage{textcomp}
\input{./sections/header}

\begin{document}
\title{Vehicle Sequencing at Signal-Free Intersections: Analytical Performance Guarantees Based on PDMP Formulation}

\author{Xiangchen Cheng, Wei Tang, Ming Yang and Li Jin
\thanks{This work was in part supported by NSFC Project 62103260, SJTU UM Joint Institute and J. Wu \& J. Sun Endowment Fund.
(Corresponding author: Li Jin.)}
\thanks{C. Cheng, M. Yang and L. Jin are with the UM Joint Institute, and W. Tang, M. Yang and L. Jin are (also) with the Department of Automation, Shanghai Jiao Tong University, China (e-mails: mercer@sjtu.edu.cn,
donquixote@sjtu.edu.cn,  mingyang@sjtu.edu.cn, li.jin@sjtu.edu.cn).}%
}

\maketitle
%=======================================================================================================================
%==================================================

\begin{abstract}
Signal-free intersections are a representative application of smart and connected vehicle technologies. Although extensive results have been developed for trajectory planning and autonomous driving, the formulation and evaluation of vehicle sequencing have not been well understood.
In this paper, we consider theoretical guarantees of macroscopic performance (i.e., capacity and delay) of typical sequencing policies at signal-free intersections. We model intersection traffic as a piecewise-deterministic Markov process (PDMP). We analytically characterize the intersection capacity regions and provide upper bounds on travel delay under three typical policies, viz. first-in-first-out, min-switchover, and longer-queue-first. We obtain these results by constructing policy-specific Lyapunov functions and computing mean drift of the PDMP. We also validate the results via a series of micro-simulation-based experiments.
\end{abstract}


\input{./sections/10_introduction} %1.5p
\input{./sections/20_model} %2.5p
\input{./sections/30_performance} %5p
\input{./sections/35_validation} %2p
\input{./sections/40_conclusion}

\bibliographystyle{IEEEtran}
\bibliography{References}   %1.5p

\end{document}

%% file: sections/header.tex
\usepackage{amsmath}

\usepackage{graphicx}
\usepackage{mathtools}
\DeclarePairedDelimiter{\ceil}{\lceil}{\rceil}

\usepackage{graphicx}

\usepackage{multirow}
\usepackage{mdwlist}
\newtheorem{theorem}{Theorem}

\usepackage{threeparttable}

\usepackage{amsfonts}
\usepackage{amsmath}
\usepackage{hyperref}

\usepackage{float}
\usepackage{caption}

\usepackage{amssymb}
\usepackage{mathrsfs}
\usepackage{subfigure}

\usepackage[ruled]{algorithm2e}
\usepackage{makecell}

%% file: sections/10_introduction.tex
\section{Introduction}
\label{sec:introduction}

Signal-free (or unsignalized) intersections are a novel traffic management system that utilizes the recent development in connected and autonomous vehicle (CAV) technology to improve capacity and abate delay \cite{pourmehrab2017,guanetti2018control, bian2019cooperation, campisi2021development}. 
The key characteristic of such systems is that vehicles are discharged as discrete ``customers'' as opposed to traffic flows at conventional signalized intersections.
Hence, signal-free intersections are more flexible and thus potentially more efficient than signalized intersections \cite{lioris2017platoons,ZHANG2022103503,wang2021roadrunner}.

\begin{figure}[hbt!]
    \centering
    \includegraphics[width=0.49\textwidth]{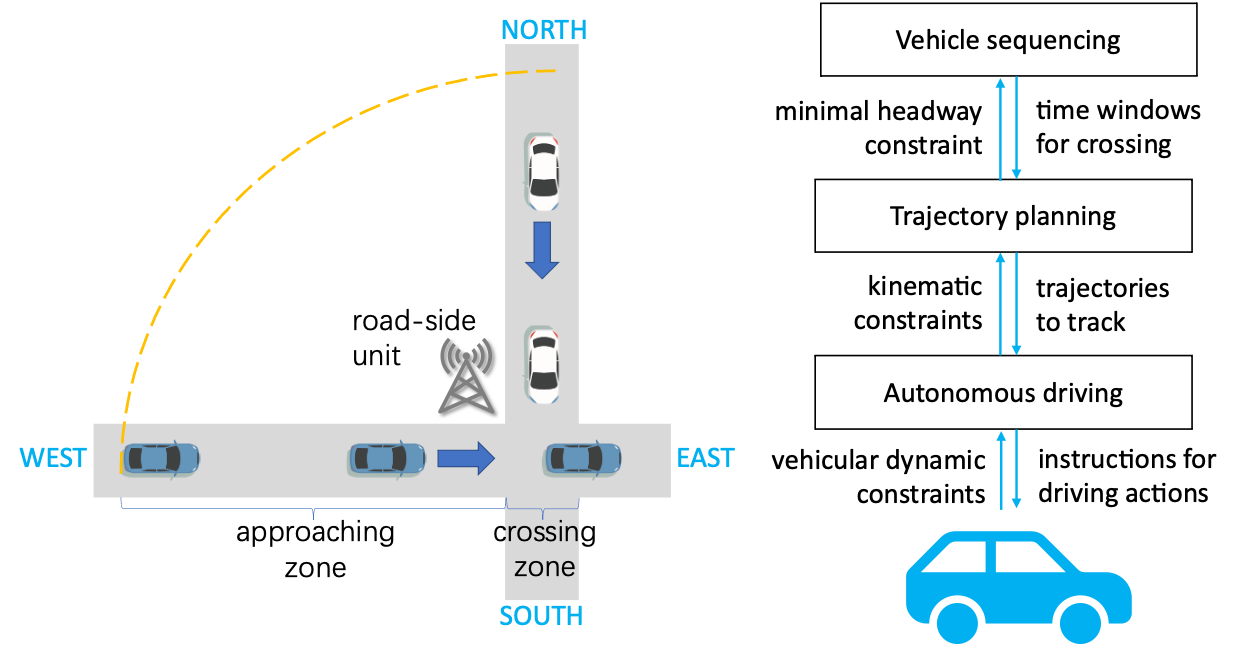}
    \caption{A two-OD intersection for CAVs and the hierarchy for decision making. This paper focuses on vehicle sequencing.}
    \label{fig:intersection}
\end{figure}

A typical signal-free intersection relies on a hierarchical decision-making mechanism as shown in Fig.~\ref{fig:intersection}  \cite{varaiya1993smart,zhang2015state,2018A}.
In this hierarchy, an infrastructure-based decision maker (e.g., a road-side unit installed at the intersection) integrates kinematic information of incoming vehicles and allocates time windows for crossing (``vehicle sequencing'').
The decisions from vehicle sequencing specify the boundary conditions for vehicles to cooperatively plan their trajectories, typically in a fully or partially centralized manner (``trajectory planning'').
The planned trajectories will be implemented by autonomous driving capabilities installed onboard (``autonomous driving'').
Upper layers instruct lower layers, while lower layers constrain upper layers.
Extensive results have been developed recently for the planning layer \cite{kim2014mpc,ahn2017safety,xu2018distributed, li2018near,2018A,mirheli2019consensus,yan2022unified} and driving layer \cite{kong2015kinematic,besselink2017string,kiran2021deep,huang2019data,wabersich2021probabilistic,cichella2020optimal}. There also exists a body of work on the sequencing layer \cite{zhu2015,ZHANG2022103503,miculescu2019polling}, but, to the best of our knowledge, very limited results are available for macroscopic evaluation of various sequencing policies.
Specifically, the following two questions have not been well understood from a theoretical, quantitative perspective:
\begin{enumerate}
    \item How many vehicles on average can the intersection discharge per hour? (Capacity)
    \item How much delay on average is induced at the intersection? (Delay)
\end{enumerate}
These two questions are essential in that we need such macroscopic performance metrics (i) to quantify the benefits of deploying relevant technologies at intersections and (ii) to compare various sequencing algorithms.

%outline & contributions
In this paper, we study the above questions by considering the sequencing problem at signal-free intersections, i.e., determining the order and times (or time windows) for incoming vehicles to cross.
We model the approaching and crossing of vehicles as a
piecewise-deterministic Markov process (PDMP) \cite{davis1984piecewise}. 
We use the Foster-Lyapunov stability theory for Markov processes to quantify key performance metrics, viz. the capacity and delay, attained by typical sequencing policies.
In particular, we identify the set of demand patterns that a certain sequencing policy can accommodate, which characterizes intersection capacity, and estimate the upper bound for delay.
We also implement the PDMP-based policies and validate the results in a standard micro-simulation platform (Simulation of Urban Mobility, SUMO \cite{krajzewicz2010traffic}). 
% \begin{figure}[H]
%     \centering
%     \includegraphics[width=0.3\textwidth]{images/intersection.png}
%     \caption{A two-orbit intersection with CAVs.}
%     \label{fig:intersection}
% \end{figure}

Sequencing is a critical decision that significantly influences intersection efficiency \cite{xu2018distributed, li2018near}, and some related work is as follows. Linear or integer programming-based sequencing algorithms have been studied in \cite{zhu2015,fayazi2018}. Lioris et al \cite{lioris2017platoons} used a classical queuing model to estimate intersection capacity with the introduction of CAVs. Zhang et al \cite{ZHANG2022103503} simulated and evaluated various sequencing policies in typical scenarios. Miculescu and Karaman \cite{miculescu2019polling} proposed an online algorithm that provides guarantees on safety and efficiency under the first-in-first-out (FIFO) policy.
In addition, the rich body of work on adaptive signalized intersections \cite{dresner2004,li2017recasting} also provide insights for the signal-free setting.
Although the above work provides very useful hints for our problem, they do not directly address the question that we consider in this paper: 

\emph{How can we analytically relate various sequencing policies to key macroscopic performance metrics?}\\
Surprisingly, to the best our knowledge, the above question has not been well understood, possibly due to lack of tractable models. This gap prohibits quantification of the macroscopic benefits of signal-free intersections (with respect to conventional intersections) and comparison between various sequencing policies.

%modeling approach
In response to the above gap, we model the traffic at an intersection as a hybrid-state PDMP. 
Every vehicle belongs to a particular class that captures the origin-destination (OD) information.
This paper focuses on the two-OD configuration as shown in Fig.~\ref{fig:intersection}, but the models, methods, and results also provide insights for multi-OD configurations.
Vehicles enter the approaching zone (Fig.~\ref{fig:intersection}) as class-specific Poisson processes and the subsequent waiting and crossing behavior is deterministic.
We define the continuous (resp. discrete) state of the PDMP as the residual system times (resp. OD classes) of all vehicles currently in the system (i.e., in the approaching or the crossing zone).
The minimal time interval between two consecutive crossings depends on the ODs of the two vehicles. The crossing times are independent and identically distributed random variables with a finite variance.
% Auxiliary state variables, which are typically discrete and sequence-dependent, may be needed to formulate certain sequencing policies.
The resultant hybrid-state PDMP model is seemingly analogous to but fundamentally different from classical queuing models \cite{sundarapandian2009probability}: In the former, the sequencing decision affects the crossing times, while in the latter, the service times are independent of service sequence.

%analysis approach
We then use the PDMP model to study key performance metrics associated with three typical sequencing policies: first-in-first-out (FIFO), min-switchover (MS), and longer-queue-first (LQF).
FIFO is the baseline policy \cite{au2010motion}, while
MS and LQF largely resemble the passing group-based policy \cite{yan2012new} and the max-pressure policy \cite{varaiya2013max}, respectively.
The main result (Theorem~\ref{thm_two}) gives criteria for traffic queue stability under various sequencing policies and, if stable, upper bounds for travel delay.
The proof is based on the Foster-Lyapunov stability theory for Markov processes \cite{meyn1993stability,benaim15,cloez15}; this is a generic theory that is conceptually important but does not directly leads to practical results in our setting.
To address this gap, we develop sequencing policy-specific Lyapunov functions that captures behavior peculiar to each policy.
For FIFO, we consider a quadratic Lyapunov function with a switching first-order term.
For MS, we construct a decomposed process that is coupled with the original PDMP but easier to analyze.
Our stability criteria are sharp (i.e., ``if and only if'') for FIFO and MS.
For LQF, we utilize peculiar properties of crossing sequences under this policy to compute the mean drift of a quadratic Lyapunov function, which leads to a sufficient condition for stability.
PDMP-based approaches have been used to study macroscopic traffic control \cite{jin2018analysis,jin2018stability}, and this paper is, to the best of our knowledge, among the first that applies such approaches to intersection control.
Therefore, our techniques themselves also contribute to the theory of traffic control.

The main result directly responds to the questions posed at the beginning of this section. The stability criteria lead to closed-form characterization for intersection capacity regions in the demand space. Using this result, we show that the capacity is the lowest if traffic is evenly distributed over various ODs, which agrees with simulation-based analysis \cite{kamal2013}. In addition, the main result provides upper bounds on the travel delays associated with various policies.
Among the three typical policies, MS attains the highest capacity and the lowest travel delay; this finding is largely consistent with simulation analysis \cite{meng2017analysis,xu2021comparison}. LQF, although providing good fairness, leads to the worst capacity and travel delay.

Finally, we discuss how the PDMP-based decisions can be translated to practically implementable instructions for CAVs and validate the results via simulation-based experiments. In particular, we show that the theoretical capacity regions adequately characterizes the simulated boundary between free flow and congestion. We also show that the theoretical upper bounds for delay are valid for the simulated values.

%Contribution
The main contributions of this paper include:
\begin{enumerate}
    \item A PDMP model of signal-free intersection that can be used to quantitatively analyze capacity and travel delay under various sequencing policies.
    \item A Foster-Lyapunov drift approach that can be used to analyze the above mentioned PDMP model in regard to macroscopic properties.
    \item Analytical characterization of capacity regions associated with typical sequencing policies.
    \item Algorithms for implementing the PDMP-based decisions and insights for vehicle sequencing.
\end{enumerate}

%paper structure
The rest of this paper is organized as follows. Section \ref{sec_model} introduces the PDMP model for intersections and formulates the sequencing policies. Section \ref{sec_performance} analyzes the theoretical properties of the above policies. Section \ref{sec_simulation} validates the theoretical results with simulations on SUMO. Section \ref{sec_conclusion} gives the concluding remarks.

%% file: sections/20_model.tex
\section{Modeling and Formulation}
\label{sec_model}

\begin{figure*}[hbt]
    \centering
    \subfigure[Actual setting.]{\includegraphics[width=0.3\textwidth]{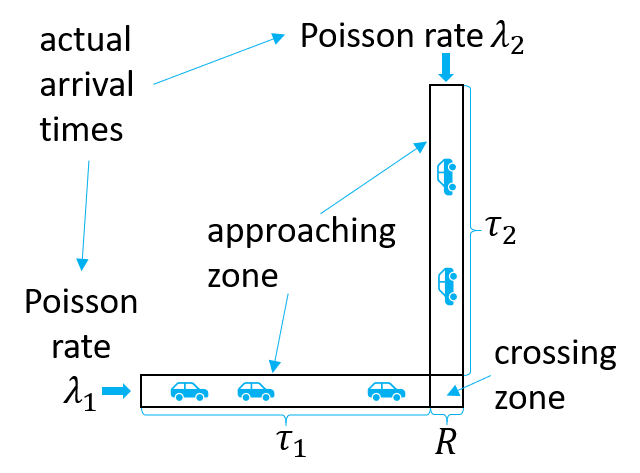}
    \label{actual}}
    \subfigure[PDMP setting.]{\includegraphics[width=0.2\textwidth]{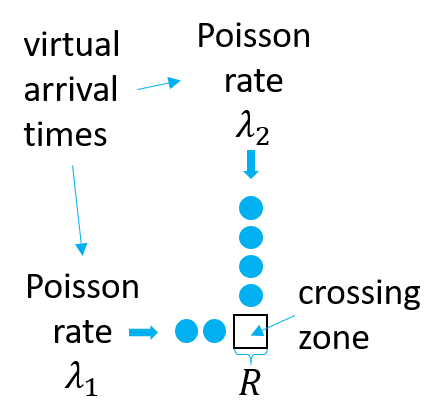}
    \label{pdmp}}
    \subfigure[PDMP state variables.]{\includegraphics[width=0.4\textwidth]{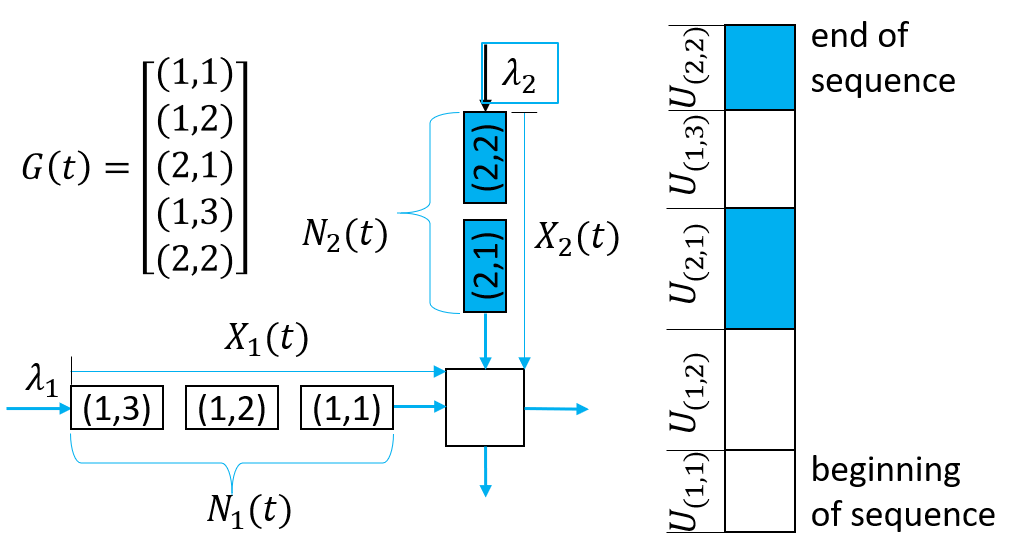}
    \label{fig:state}}
    \caption{Modeling intersection as PDMP.}
    \label{fig:two_config}
\end{figure*}

In this section, we formulate the sequencing problem for the intersection shown in Fig.~\ref{actual}.
We first specify the PDMP model. Then, we formulate the sequencing policies that we will study. Finally, we define the performance metrics for evaluation and comparison.

%%%%%%%%%%%%%%%%%%%%%%%%%%%%%%%%%%%%%%
\subsection{Piecewise-deterministic Markov model}

We consider random arrivals.
For the two-OD intersection in Fig.~\ref{actual}, let $\mathcal K=\{1,2\}$ be the set of OD pairs (or traffic classes). Class-$k$ vehicles arrive at the approaching zone as the Poisson process of rate $\lambda_k$.
If a class-$k$ vehicle arrives and sees no other vehicles (of either class), then the vehicle will traverse the approaching zone with a nominal traverse time $\tau_k$. Suppose that a class-$k$ vehicle arrives at the approaching zone at time $t_0$, which we call the \emph{actual arrival time}; then, we define $t_1=t_0+\tau_k$ as the \emph{virtual arrival time}, or simply \emph{arrival time}, of the vehicle. That is, if this vehicle were not to be influenced by other vehicles, it would arrive at the crossing zone at time $t_1$. Since the nominal traverse time is constant, if one observes at the crossing zone, the arrival process for class $k$ is also Poisson of rate $\lambda_k$. Hence, instead of the actual kinematics of incoming vehicles, we consider a PDMP formulation for the equivalent, virtual ``queuing process'', where the arrival times are equal to the above-mentioned virtual arrival times; see Fig.~\ref{pdmp}. In Section~\ref{sec_simulation}, we will discuss how sequencing policies formulated in the PDMP can be translated back to instructions implementable in the approaching zone.

\begin{figure}[hbt]
    \centering
    \includegraphics[width=0.45\textwidth]{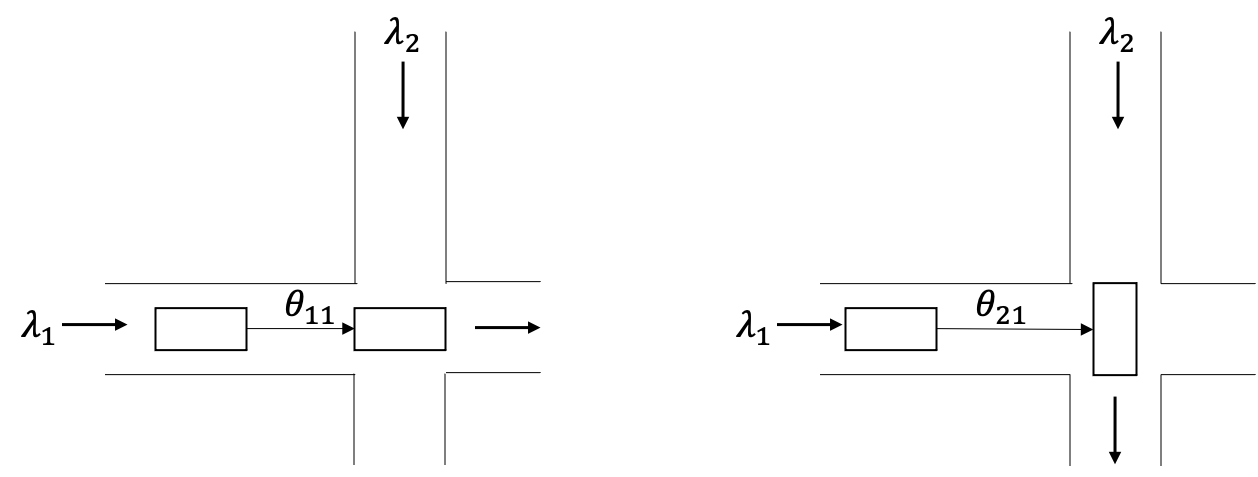}
    \caption{Minimal headway depends on the sequence.}
    \label{fig:two_theta}
\end{figure}

The evolution of the PDMP is driven by the arrival and discharge of vehicles. When a class-$k$ vehicle arrives at the crossing zone at time $t$ and sees no other vehicle crossing, then it will finish crossing at time $t+R$, where $R$ is the \emph{crossing time} (with unit [sec]).
To account for vehicle heterogeneity, we assume that $R$ is a random variable (rv) with a cumulative distribution function $F_R(r)$ supported by a bounded interval $[R_{\min},R_{\max}]$; the value of $R$ becomes known to the system operator when a vehicle arrives. In practice, $R$ depends on vehicle size and crossing speed. Since $R$ is bounded, so are the mean $\bar R$, the variance $\sigma_R^2$, and the moment generating function (MGF) $g_R(\rho)$.
If an incoming vehicle sees other vehicles crossing or waiting for crossing, it will have to wait. Note that such queuing-like waiting is virtual in the PDMP; in practice, the waiting time will be absorbed over the approaching zone via trajectory planning; see Section~\ref{sub:two_sim}.
For any two vehicles crossing the intersection consecutively, the headway (with unit [sec]) in between must be no less than the \emph{minimal headway} $\theta_{ij}\ge0$, where $i$ (resp. $j$) are the class of the leading (resp. following) vehicle. 
Thus, we have a headway matrix $\Theta\in\mathbb R_{\ge0}^{2\times2}$. By practical insight, we assume that $\theta_{ij}>\theta_{ii}$ for $j\ne i$ and $i\in\mathcal K$; see Fig.~\ref{fig:two_theta}.
Since the \emph{service times} $\theta_{ij}+R$ are not independent and identically distributed, classical queuing theory does not apply to our PDMP here, especially for the purpose of sequencing policy analysis; new tools, which this paper focuses on, are needed.

A complete state-space representation of the PDMP is as follows.
Let $N_k(t)\in\mathbb Z_{\ge0}$ be the number of class-$k$ vehicles waiting for discharge at time $t$, and let $N(t)=N_1(t)+N_2(t)$; see Fig.~\ref{fig:state}. We call $N_k(t)$ the \emph{class-$k$ count} and $N(t)$ the \emph{total count}. We use a tuple $(k,n)$ to label the $n$th class-$k$ vehicle. Vehicles will cross the intersection according to the \emph{sequence of crossing} $G(t)=[G_1(t),G_2(t),\ldots,G_{N(t)}]^T\in(\mathcal K\times\mathbb Z_{>0})^{N(t)}$, where $G_i(t)=(k,n)$ means that, observed at time $t$, vehicle $(k,n)$ will be the $i$th to cross the intersection if no more vehicles are joining the queues.
Let $U_{k,n}(t)$ be the \emph{residual service time} for vehicle $(k,n)$ defined as follows. If $(k,n)\ne G_1(t)$, i.e., if vehicle $(n,k)$ is not at the beginning of the sequence, then $U_{k,n}(t)$ is the service time for vehicle $(n,k)$. If $(k,n)=G_1(t)$, then vehicle $(k,n)$ is ``being served'' and will fully cross the intersection at time $t+U_{k,n}(t)$. Let $U(t)$ be the vector of $U_{k,n}(t)$ for $1\le n\le N_k(t)$ and $k\in\mathcal K$. Thus, the evolution of the PDMP can be fully described by the hybrid state $(G(t),U(t))$, where $G(t)$ is discrete and $U(t)$ is continuous. Since the vehicle arrivals are Poisson, if the sequencing policy is also Markovian (i.e., depending on $G(t)$ and $U(t)$ only), $\{(G(t),U(t));t\ge0\}$ is a piecewise-deterministic Markov process \cite{davis1984piecewise}.
% the state space is $\mathcal G\times\mathcal U$, where $\mathcal G=\cup_{n\in\mathbb Z_{\ge0}}(\mathcal K^n\times\mathbb Z_{>0}^n)$ is the discrete state space and $\mathcal U=\cup_{n\in\mathbb Z_{\ge0}}\mathbb R_{\ge0}^n$ is the continuous state space.

For ease of presentation, we also introduce a simplified representation.
Although $G(t)$ and $U(t)$ give a complete state-space representation, they have time-varying dimensions and are thus not easy to use directly.
To resolve this problem, we define a an \emph{aggregate state} $X(t)=[X_1(t),X_2(t)]^T$, where
$$
X_k(t)=\sum_{n=1}^{N_k(t)}U_{k,n}(t),
\quad k=1,2.
$$
If $N_k(t)=0$, we define the above to be zero.
One can interpret $X_k(t)$ as the ``aggregate service time'' or ``temporal queue size'' of class-$k$ traffic.

Finally, the dynamics for the hybrid state $(G(t),U(t))$ and thus the aggregate state $X(t)$ depends on the sequencing policy, which we discuss in the next subsection.

%%%%%%%%%%%%%%%%%%%%%%%%%%%%%%%%%%%%%%
\subsection{Sequencing policies}

The PDMP formulation is incomplete until a sequencing policy is specified. Intuitively, the sequencing policy defines how the hybrid state $(G(t),U(t))$ or the aggregate state $X(t)$ is reset when a vehicle arrives. For ease of presentation, we introduce sequencing policies in terms of the aggregate state $X(t)$.
Between vehicle arrivals, $X(t)$ evolves deterministically as follows:
\begin{align*}
    \frac d{dt}X_k(t)=\begin{cases}
    -1 & \mbox{$(k,1)$ is being discharged,}\\
    0 & \mbox{otherwise,}
    \end{cases}
    \quad k=1,2.
\end{align*}
Arrivals of vehicles will lead to sudden jumps in $X(t)$. The magnitude of such jumps depends on the sequencing policy.
We consider three typical sequencing policies:
\begin{enumerate}
    \item {\bf First-in-first-out (FIFO)} discharges vehicles according to the order of arrivals. To formulate this policy, we define an auxiliary discrete state $Y(t)\in\mathcal K$
    that tracks the class of the vehicle at the end of the sequence: we define $Y(t)=k$ if a class-$k$ vehicle is at the end of the sequence; if $N(t)=0$, then $Y(t)$ is the class of the last discharged vehicle. Thus, $\{Y(t);t\ge0\}$ is a two-state Markov process with transition rates $q_{12}=\lambda_1$ and $q_{21}=\lambda_2$. If a class-$k$ vehicle arrives at time $t$, the continuous state $X(t)$ is then updated as follows:
    \begin{align*}
        &X_k(t)=
        X_k(t_-)+\theta_{Y(t_-),k}+R,\\
        &X_{-k}(t)=X_{-k}(t_-),
    \end{align*}
    where ``$-k$'' means $j\in\mathcal K:j\ne k$ and ``$t_-$'' means left limits, for Poisson processes are right-continuous with left limits (RCLL) \cite{gallager2013stochastic}.
    % The discrete state $Y(t)$ evolves according to the mechanism in Fig.~\ref{fig:two_Yt}.
    % \begin{figure}[hbt]
    %     \centering
    %     \includegraphics[width=0.5\textwidth]{images/two_Yt.png}
    %     \caption{Transition mechanisms for discrete states $Y(t)$ (left) and $Z(t)$ (right). $\lambda_k$ indicates the rate of Poisson transitions, while equalities/inequalities indicate the condition for deterministic transitions.}
    %     \label{fig:two_Yt}
    % \end{figure}
    
    \item {\bf Min-switch (MS)} always clears traffic in one class before switching to the other class. To formulate this policy, we need an auxiliary discrete state $Z(t)\in\mathcal K$ that tracks the class being discharged: we define $Z(t)=k$ if a class-$k$ vehicle is crossing or was the last to cross the intersection. If a class-$k$ vehicle arrives at time $t$, the continuous state $X(t)$ is then updated as follows:
    \begin{align*}
        &X_k(t)=\begin{cases}
        X_k(t_-)+\theta_{k,k}+R,&X(t_-)>0,\\
        \theta_{-k,k}+R & \mbox{otherwise,}
        \end{cases}\\
        &X_{-k}(t)=X_{-k}(t_-).
    \end{align*}
    The discrete state $Z(t)$ is updated only when traffic in class $Z(t_-)$ has been cleared and when there is non-zero traffic in class $-Z(t_-)$ waiting for discharge.
    
    \item {\bf Longer-queue-first (LQF)} always discharges the class with a longer (in a generalized sense) queue. 
    This policy leads to dynamics more sophisticated than the other policies, so we will need the full state $(G(t),U(t))$.
    For ease of presentation, we use $Q(t)\in\mathcal H$ to track which OD has a longer ``queue'':
    \begin{align*}
        Q(t)=\begin{cases}
        1 & X_1(t)>\beta X_2(t),\\
        2 & X_1(t)<\beta X_1(t),\\
        0 & \mbox{otherwise,}
        \end{cases}
    \end{align*}
    where $\beta>0$ is a design parameter.
    Note that here we compare the ``temporal queue size'' $X_k(t)$ instead of the vehicle counts $N_k(t)$, since the former are more relevant for delay balancing.
    If a class-$k$ vehicle arrives at time $t$, the aggregate state $X(t)$ is then updated as follows:
    \begin{align*}
        &X_k(t)=\begin{cases}
        X_k(t_-)+\theta_{k,k}+R& Q(t_-)=k\mbox{ or }0,\\
        X_k(t_-)+\theta_{-k,k}+R & Q(t_-)=-k,
        \end{cases}\\
        &X_{-k}(t)=\begin{cases}
        X_{-k}(t_-) \hspace{1.85cm} Q(t_-)=k\mbox{ or }0,\\
        X_{-k}(t_-)+\theta_{-k,k}-\theta_{-k,-k}\\
        \hspace{3.18cm} Q(t_-)=-k,
        \end{cases}
    \end{align*}
    That is, the new arrival will be placed at the end of the sequence if it belongs to the longer queue. Otherwise, it will be inserted in the middle of the sequence before some vehicles in the longer queue; see Fig.~\ref{fig:two_etat}.
    \begin{figure}[hbt]
        \centering
        \includegraphics[width=0.45\textwidth]{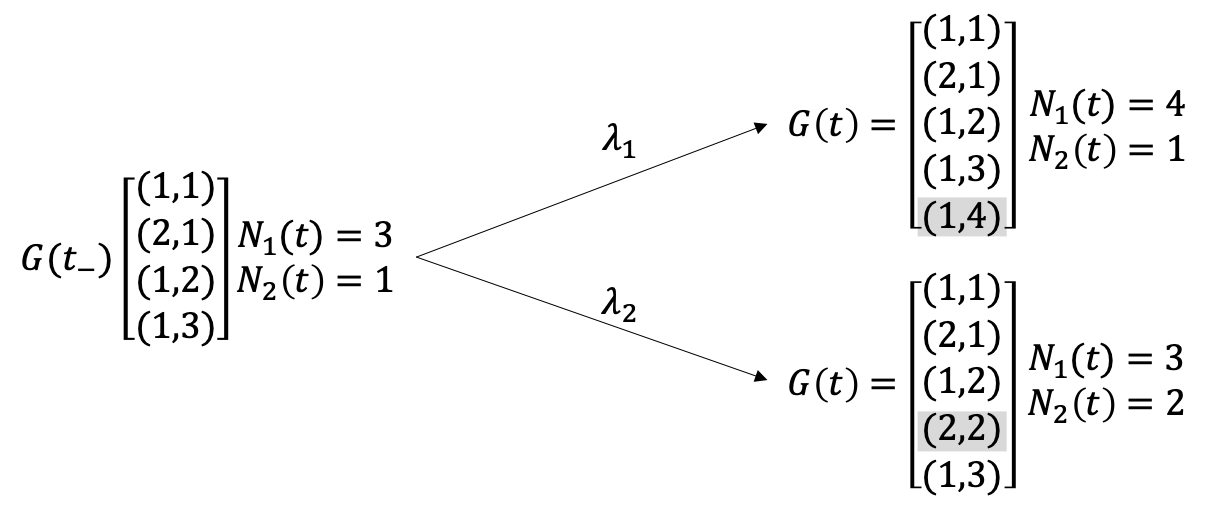}
        \caption{LQF policy may sequence new arrivals (shaded) before existing vehicles to balance traffic from different directions.}
        \label{fig:two_etat}
    \end{figure}
    % Note that $(Q(t),X(t))$ is not a complete state-space representation and is used only as auxiliary states.
\end{enumerate}

\begin{figure}[hbt]
    \centering
    \includegraphics[width=0.3\textwidth]{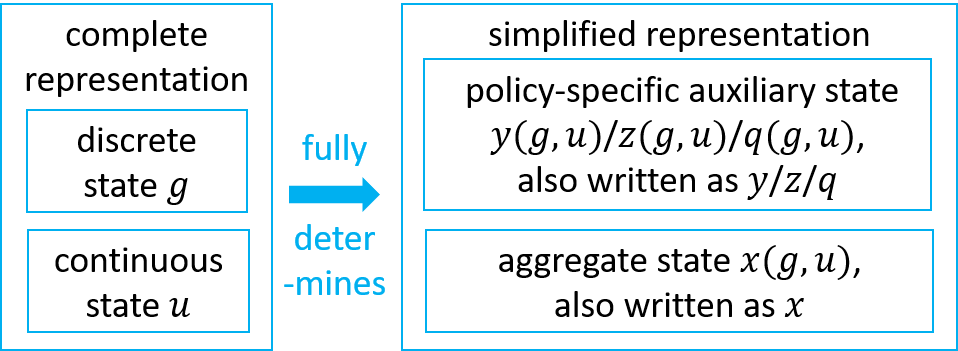}
    \caption{Two representations for PDMP.}
    \label{representation}
\end{figure}

Note that all auxiliary states in the above can be uniquely derived from the PDMP's complete state $(G(t),U(t))$; see Fig.~\ref{representation}.
Hence, given a sequencing policy, $\{(G(t),U(t));t\ge0\}$ is a PDMP. One can verify that it is RCLL.
Thus, the PDMP dynamics can be compactly characterized with the \emph{generator} $\mathscr A$ defined as follows.
Let $V:\mathcal G\times\mathcal U\to\mathbb R$ be a well-defined (in the sense of \cite[p.521]{meyn1993stability}) function.
Then, $\mathscr AV$ is essentially a measurable function such that for each initial condition $(g,u)\in\mathcal G\times\mathcal U$ and each time $t>0$,
\begin{align*}
    &\mathrm E\bigg[V\Big(G(t),U(t)\Big)\bigg|\Big(G(0),U(0)\Big)=(g,u)\bigg]=V(g,u)\\
    &+\mathrm E\bigg[\int_{s=0}^t\mathscr AV\Big(G(s),U(s)\Big)ds\bigg|\Big(G(0),U(0)\Big)=(g,u)\bigg],\\
    &\mathrm E\bigg[\int_{s=0}^t\mathscr AV\Big(G(s),U(s)\Big)ds\bigg|\Big(G(0),U(0)\Big)=(g,u)\bigg]<\infty.
\end{align*}
The term $\mathscr AV(g,u)$ is typically interpreted as the time derivative of $\mathrm E[V(G(t),U(t))]$ as $(G(t),U(t))=(g,u)$. Hence, $\mathscr AV(g,u)$ is called the \emph{mean drift} of $V$. Stability of the PDMP is closely related to behavior of the mean drift of appropriate Lyapunov functions \cite{meyn1993stability,benaim15,cloez15}.

% The PDMP abstracts practical traffic dynamics at two-OD intersections. In Sections~\ref{sub:two_sim} and \ref{sub:two_exp}, we will show how outputs of the PDMPs can be translated to instructions that can be implemented in practice.

%%%%%%%%%%%%%%%%%%%%%%%%%%%%%%%%%%%%%%
\subsection{Performance metrics}

To obtain analytical performance guarantees (delay and capacity) using the PDMP, we consider the long-time behavior of the aggregate state $X(t)$. We say that the PDMP is \emph{stable} if there exists $\overline{W}<\infty$ such that for any initial condition,
\begin{align}
    \limsup_{t\to\infty}\frac1t\int_{s=0}^t\mathrm E[\|X(s)\|_1]ds\le\overline{W},
    \label{eq_zeta}
\end{align}
where $\|\cdot\|_1$ is the 1-norm in Euclidean spaces.
Furthermore, if the above holds, there typically exists $\overline X<\infty$ such that for any initial condition
\begin{align}
    \lim_{t\to\infty}\frac1t\int_{s=0}^t \|X(s)\|_1ds=\overline X
\quad a.s.;
\label{eq_Xbar}
\end{align}
see Section~\ref{sub_fifo} for discussion on the existence of this limit.

Practically, $\overline X$ can be interpreted as the average \emph{delay} due to vehicle coordination. Computing the exact value of $\overline X$ is in general not easy, so we focus on obtaining an upper bound $\overline{W}$. Stability in the sense of \eqref{eq_zeta} means the delay is within an acceptable range, while instability means congestion will grow unboundedly. We will compare the delays resulting from various sequencing policies.

The \emph{capacity} of the intersection is essentially the maximal demand that can be accommodated in the sense that \eqref{eq_zeta} holds, i.e., that the PDMP is stable. Since the demand is a two-dimensional vector $\lambda=[\lambda_1,\lambda_2]^T$, we cannot use a scalar to characterize the intersection's capacity. Instead, we are interested in identifying the set of $\lambda$ such that the PDMP is stable, which is called the \emph{capacity region}. Note that various sequencing policies lead to different capacity regions, which enables us to compare policies in terms of capacity.

In particular, if we fix the \emph{distribution of demand}, then we can define capacity as a scalar as follows. 
The distribution of demand is characterized by the probability vector $p=[p_1,p_2]$.
With a fixed $p$, we can define capacity $\bar\lambda$ as follows:
\begin{align*}
    \bar\lambda=\max&\quad \|\lambda\|_1\\
    s.t.&\quad \lambda=p\|\lambda\|_1,\ \lambda\ge0,\\
    &\quad \mbox{stability condition holds.}
\end{align*}

The next section studies the analytical guarantees of the above performance metrics.

%% file: sections/30_performance.tex
%%%%%%%%%%%%%%%%%%%%%%%%%%%%
\section{Analytical performance guarantees}
\label{sec_performance}

In this section, we derive analytical guarantees on performance metrics under various sequencing policies.
Recall questions 1 and 2 in Section~\ref{sec:introduction}, which were posed in the practical setting. These questions can be mapped to the following questions in the PDMP setting:
\begin{enumerate}
    \item When does the upper bound in \eqref{eq_zeta} exist? (Capacity analysis.)
    \item If an upper bound exists, how can we estimate it? (Delay analysis.)
\end{enumerate}
This section is devoted to answering the above questions and deriving practical insights.

To state the main result, let $\|\cdot\|_p$ be the $p$-norm and $J_{m,n}$ be the $m\times n$ matrix of ones. Let $\mathrm{diag}(\cdot)$ be the row vector of the diagonal components of a matrix. Let $\circ$ be the Hadamard product operator. Let $\mathrm{det}(\cdot)$ be the determinant of a matrix. Define
{
\begin{subequations}
\begin{align}
    &
    A(\lambda)=\nonumber\\
    &\begin{bmatrix}
    $$
    \frac{\lambda_1(\theta_{11}-\theta_{21})+\lambda_2(\theta_{12}-\theta_{22})}{2(\lambda_1+\lambda_2)}
    &
    \frac{\lambda_1(\theta_{11}-\theta_{21})+\lambda_2(\theta_{12}-\theta_{22})}{2(\lambda_1+\lambda_2)}
    \\
    \frac{\lambda_1(\theta_{21}-\theta_{11})+\lambda_2(\theta_{22}-\theta_{12})}{2(\lambda_1+\lambda_2)}
    &
    \frac{\lambda_1(\theta_{21}-\theta_{11})+\lambda_2(\theta_{22}-\theta_{12})}{2(\lambda_1+\lambda_2)}
    $$
    \end{bmatrix},
    \\
    &B(\lambda)=[b_{ij}(\lambda)]\nonumber\\
    &{\scriptsize\begin{bmatrix}
$$
(\theta_{12}+\theta_{21}+\overline{R}+R_{\max})\lambda_1
&
(\theta_{11}+\overline{R})\lambda_1+(\theta_{21}-\theta_{11})\lambda_2-1
\\
(\theta_{12}-\theta_{22})\lambda_1+(\theta_{22}+\overline{R})\lambda_2-1
&
(\theta_{12}+\theta_{21}+\overline{R}+R_{\max})\lambda_1
$$
\end{bmatrix}}.
\label{eq_B}
% \\
% &\gamma(\lambda)=\frac{\max\Big\{b_{11}(\lambda)+\beta b_{12}(\lambda),\beta b_{21}(\lambda)+b_{22}(\lambda)\Big\}}{\sqrt{1+\beta^2}}
\end{align}
\end{subequations}
}%
Then, we state the main result of this paper as follows.

\begin{theorem}
\label{thm_two}
Suppose an intersection with arrival rates $\lambda\in\mathbb R_{\ge0}^2$ and headway matrix $\Theta$. Let $\bar R$ and $\sigma_R^2$ be the mean and variance of the crossing time.
\begin{enumerate}
    \item[(i)] FIFO policy is stabilizing if and only if
    \begin{align}
        \Big(\frac{\lambda^T}{\|\lambda\|_1}\Theta+\bar RJ_{1,2}\Big)\lambda<1.
        \label{eq_fifo}
    \end{align}
    Furthermore, if \eqref{eq_fifo} holds, then
    \begin{align}
    \overline W_0
    \le
        \overline X
        \le \overline W_1,
        \label{eq_Wbar1}
    \end{align}
    where
    {\footnotesize
    \begin{align*}
        &\overline W_0=\mathrm{diag}(\Theta)\frac{\lambda}{\|\lambda\|_1}+\overline R\\
        &\quad+\frac{\Big(\mathrm{diag}(\Theta\circ\Theta)+\mathrm{diag}(\Theta)\overline{R}\Big)\frac{\lambda}{\|\lambda\|_1}+\overline R^2+\sigma_R^2}{\frac2{\|\lambda\|_1}-2\|\lambda\|_1\Big(\mathrm{diag}(\Theta)\frac{\lambda}{\|\lambda\|_1}+\overline R\Big)},
        \\
        &\overline W_1\\
        &=\frac{\Big\|\Big((\Theta+\overline RJ_{2,2})\circ(\frac12(\Theta+\overline RJ_{2,2})+A(\lambda))+\frac12\sigma_R^2J_{1,2}\Big)\lambda\Big\|_\infty}{1-\Big(\frac{\lambda^T}{\|\lambda\|_1}\Theta+\overline RJ_{1,2}\Big)\lambda}.
    \end{align*}}
    
    \item[(ii)] MS policy is stabilizing if and only if
    \begin{align}
        \mathrm{diag}(\Theta+\bar RJ_{2,2})\lambda<1.
        \label{eq_ms}
    \end{align}
    Furthermore, if \eqref{eq_ms} holds, then
    \begin{align}
    \overline W_0
    \le
        \overline X\le \overline W_2,
        \label{eq_Wbar2}
    \end{align}
    where 
    {\footnotesize\begin{align*}
    \overline W_2 =&\frac{\Big\|\Big((\Theta+\overline RJ_{2,2})\circ(\Theta+\overline RJ_{2,2})+\sigma_R^2J_{1,2}\Big)\lambda\Big\|_\infty}{2-2\mathrm{diag}(\Theta+\overline RJ_{2,2})\lambda}\\
    &+\frac{\lambda_1\lambda_2}{\lambda_1+\lambda_2}\sum_{k=1}^2(\theta_{-k,k}-\theta_{k,k}).
    \end{align*}}
    
    \item[(iii)] LQF policy is stabilizing if
    \begin{align}
        % &(\theta_{12}+\theta_{21}+\overline{R}+R_{\max})^2\lambda_1\lambda_2\nonumber
        % \\
        % &<\Big(1-(\theta_{11}+\overline{R})\lambda_1-(\theta_{21}-\theta_{11})\lambda_2\Big)\Big(1-(\theta_{12}\nonumber
        % \\
        % &\quad-\theta_{22})\lambda_1-(\theta_{22}+\overline{R})\lambda_2\Big)
        \mathrm{det}\Big(B(\lambda)\Big)<0
        \label{eq_lqf}
    \end{align}
    and if the comparing parameter $\beta$ satisfies
    \begin{align}
        \frac{b_{11}(\lambda)}{-b_{21}(\lambda)}
        <\beta
        <\frac{-b_{12}(\lambda)}{b_{22}(\lambda)}.
        \label{eq_beta}
    \end{align}
    Furthermore, if \eqref{eq_lqf}--\eqref{eq_beta} hold, then
    \begin{align}
        \overline W_0
    \le
    \overline X\le \overline W_3,
    \label{eq_Wbar3}
    \end{align}
    where
    {\footnotesize\begin{align*}
    \overline W_3 =\sqrt{\frac{(1+\beta^2)}{2}}\frac{\max_k\{b_{kk}^2(\lambda)+(b_{k,-k}(\lambda)+1)^2\}+\sigma_R^2\|\lambda\|_1}{-\max_k\{b_{kk}(\lambda)+\beta b_{k,-k}(\lambda)\}}.
    \end{align*}}
\end{enumerate}
\end{theorem}

The above theorem directly responds to the questions posed at the beginning of this section. 
The stability criteria for FIFO and MS are necessary and sufficient, while that for LQF is sufficient. Hence, we can compute the exact capacity under FIFO and MS and estimate a lower bound for capacity under LQF.
The delay lower bound $\overline{W}_0$ is common for all policies, while the upper bounds are policy-specific.

The rest of this section is devoted to an example-based interpretation (Section~\ref{sub_numerical}) and proof (Section~\ref{sub_proof}) of the theorem.

%%%%%%%%%%%%%%%%%%%%%%%%%%%%%%%%%%%%%%%%%%%%%%%%%%
\subsection{Numerical example}
\label{sub_numerical}
\begin{figure}[hbt]
    \centering
    \includegraphics[width=0.5\textwidth]{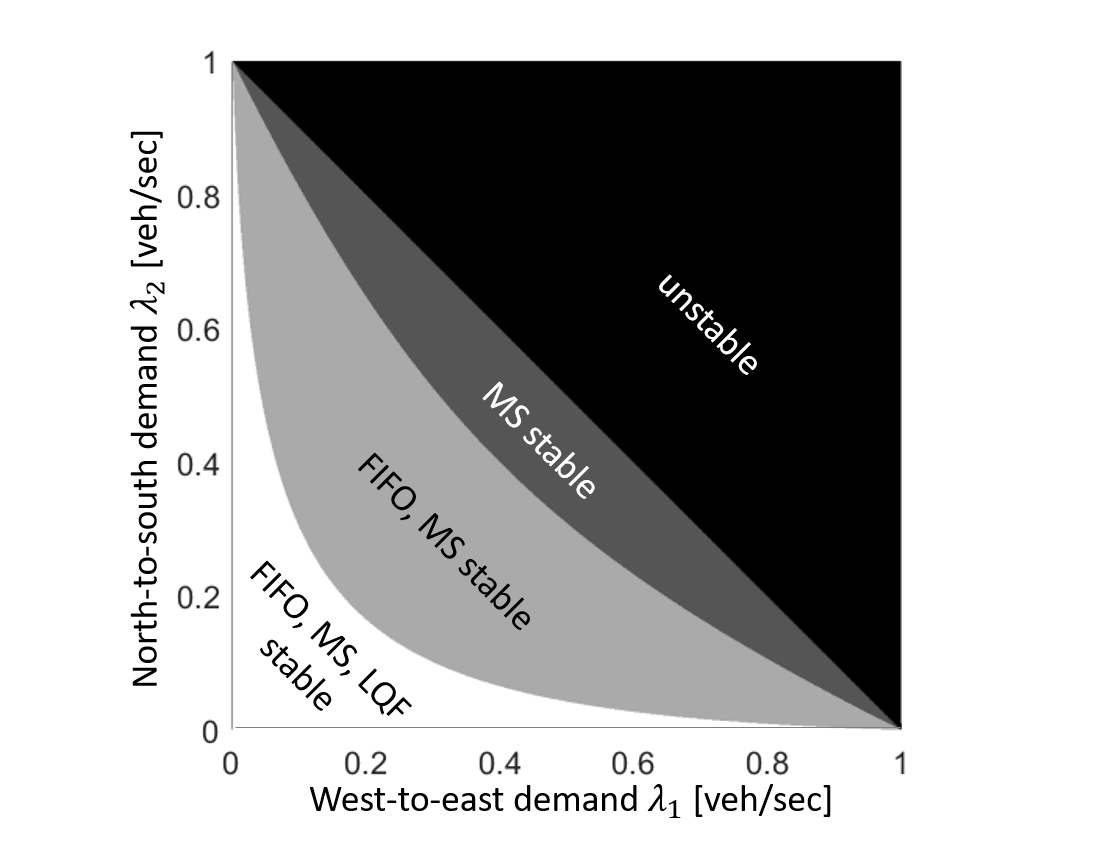}
    \caption{Stability regimes under various policies. MS stabilizes the white, light gray, dark gray regimes, FIFO stabilizes the white and light gray regimes, and LQF stabilizes the white regime only.}
    \label{fig:two_regimes}
\end{figure}

A major use of Theorem~\ref{thm_two} is to compute the capacity regions under various policies. 
Fig.~\ref{fig:two_regimes}
shows the stability regimes associated with parameters
\begin{align*}
    {\Theta} = 
\begin{bmatrix}
$$
0.5 & 1\\
1 & 0.5
$$
\end{bmatrix} \mbox{[sec]},\ 
\bar R=0.5\ \mbox{[sec]},\ 
\sigma_R^2=0.1.
\end{align*}
That is, it takes every vehicle 0.5 sec on average to go through the crossing zone, and the inter-vehicle headway should be no less than 0.5 or 1 sec, depending on the crossing sequence.
For LQF, we assume that $\beta=1$ and that class 1 is prioritized in case of a tie.

As Fig.~\ref{fig:two_regimes} shows, MS (resp. LQF) leads to the largest (resp. smallest) stable regime,  and thus the largest capacity region. Therefore, MS gives higher capacity while LQF gives lower capacity.
The reason is that switching over the direction of crossing requires additional time, and MS minimizes the chance of switchovers. LQF, on the contrary, maximizes the chance of switchovers, for this policy tends to discharge various ODs in an alternate manner so that two traffic queues are balanced. FIFO is to some extent in the middle of MS and LQF.

\begin{figure}[hbt]
    \centering
    \includegraphics[width=0.4\textwidth]{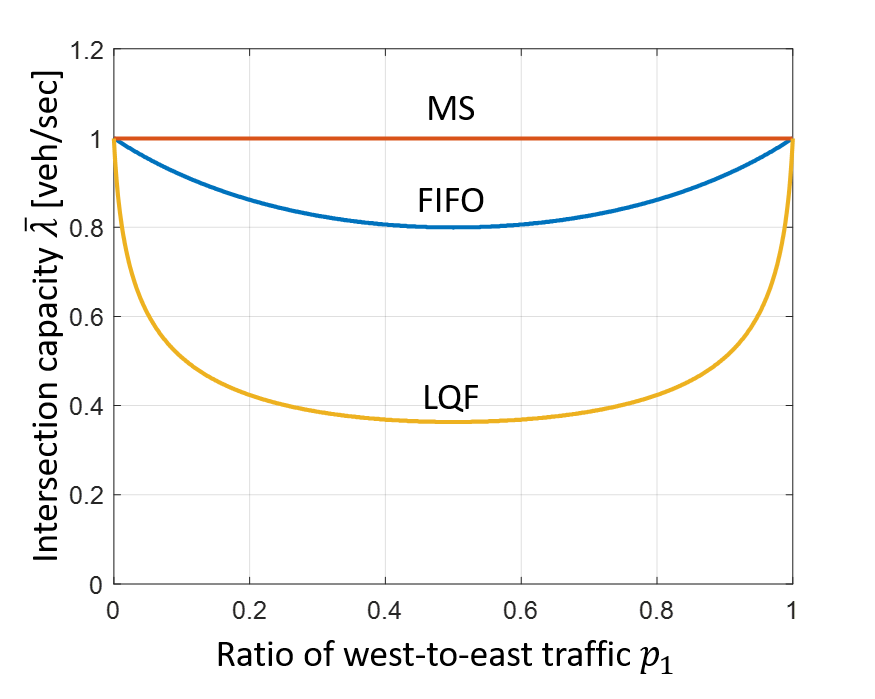}
    \caption{MS maximizes capacity for any demand distribution.}
    \label{fig:two_lambdabar}
\end{figure}

An important insight from Theorem~\ref{thm_two} is that the capacity of an intersection depends on both the distribution of demand and the sequencing policy. 
The capacity depends on the distribution of demand via $p$ and on the sequencing policy via the stability condition \eqref{eq_fifo} or \eqref{eq_ms} or \eqref{eq_lqf}. As Fig.~\ref{fig:two_lambdabar} shows,
if the demand distribution is balanced (i.e., $p_1\approx p_2$), the differences between the policy-specific capacities are large; if the demand distribution is highly imbalanced (i.e., $p_1\approx1$ or $p_2\approx1$), then such differences are minimal. The reason is that imbalanced demand makes switchovers less frequent, and thus the impact of sequencing policy is less significant.

In conclusion, as far as capacity is concerned,
Theorem~\ref{thm_two} implies the following:
\begin{enumerate}
    \item MS maximizes capacity in that given the headway matrix $\Theta$, no sequencing policy can attain a higher capacity than MS does.
    
    \item FIFO performs as well as MS if $p_1\approx1$ or $p_2\approx1$ but not so well if $p_1\approx p_2$, since $p_1\approx1$ or $p_2\approx1$ means low chance of alternate sequences and thus fewer switchovers, while $p_1\approx p_2$ means high chance of alternate sequences and thus more switchovers.
    
    \item LQF in general gives the lowest capacity, since this policy is not intended to maximize discharge rate. Instead, this policy focuses on balancing traffic with various ODs.
\end{enumerate}

\begin{figure}[hbt]
    \centering
    \includegraphics[width=0.45\textwidth]{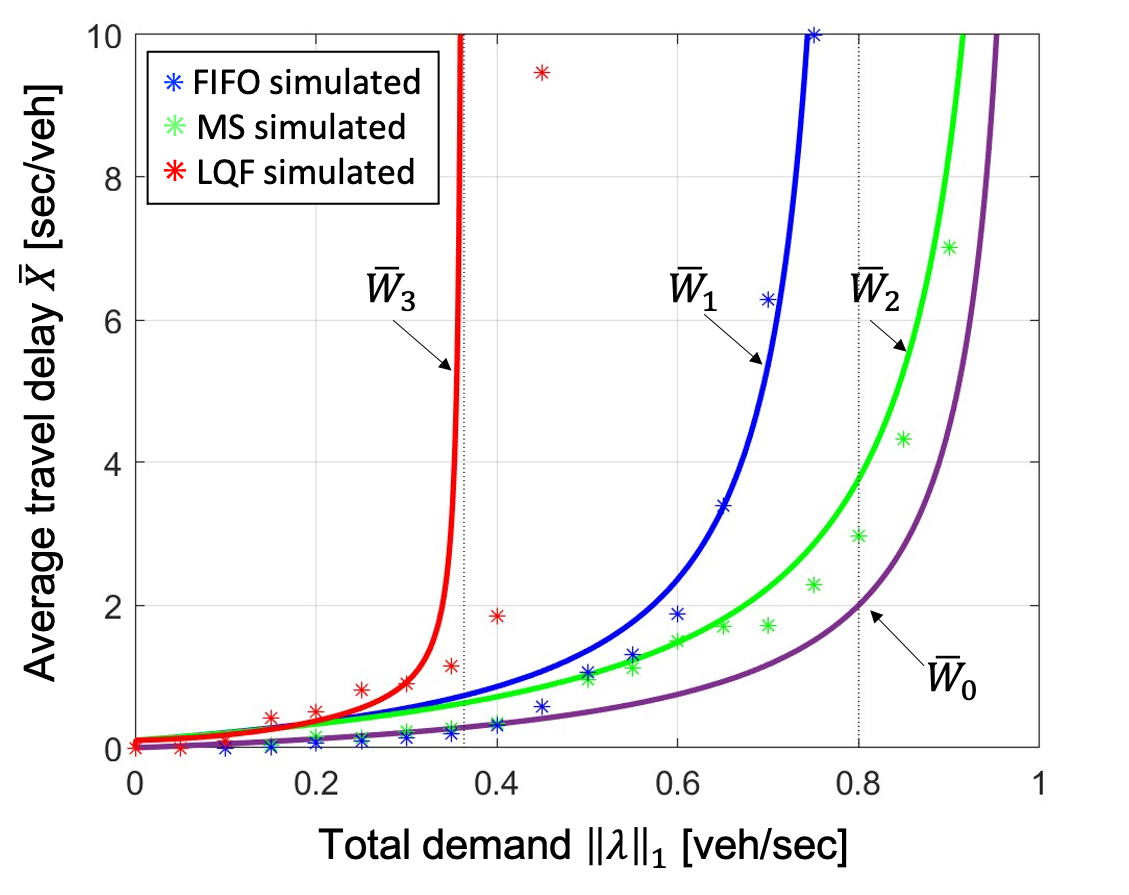}
    \caption{Simulated average delay $\overline X$ in SUMO and theoretical bounds given by Theorem~\ref{thm_two}. Symmetric demand pattern (i.e., $p_1=p_2=0.5$) is assumed.}
    \label{fig:two_Wbar}
\end{figure}

When the intersection is stabilized, we can also use the bounds provided in Theorem~\ref{thm_two} to estimate the average delay. Fig.~\ref{fig:two_Wbar} shows that MS gives not only the maximal capacity but also the minimal delay. In fact, the delay under MS ($\overline{W}_2$ in Fig.~\ref{fig:two_Wbar}) is very close to the theoretical lower bound $\overline{W}_0$, which results from an over-optimistic estimation (see Section~\ref{sub_fifo}).
LQF leads to both minimal capacity and maximal delay.

\begin{figure}[hbt]
    \centering
    \includegraphics[width=0.25\textwidth]{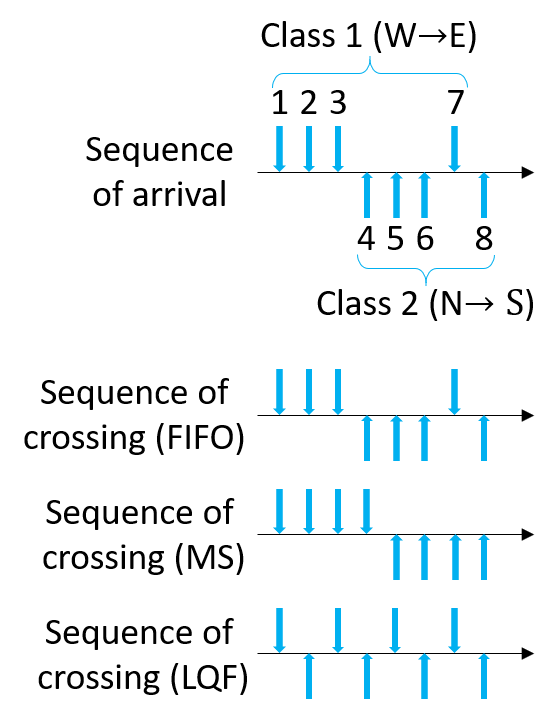}
    \caption{Sequences of arrival/crossing for a given sample path. Vehicle 1 (resp. 8) is the first (resp. last) arrival.}
    \label{fig:sequence}
\end{figure}

We now use a specific sample path to further explain the above insights. Suppose 8 vehicles already in the system with a given arrival sequence (Fig.~\ref{fig:sequence}).
We assume that no more vehicle will arrive and focus on the times for each policy to discharge these 8 vehicles. Fig.~\ref{fig:sequence} shows the sequences of crossing under each policy. Because of the switchover time, more turns mean more delay. MS minimizes the number of turns, while LQF creates the most switchovers. The total discharge times are listed in Table~\ref{tab:discharge}.
\begin{table}[hbt]
    \centering
    \begin{tabular}{|c|c c c|}
        \hline
        System time for & FIFO & MS & LQF \\
        \hline
        Last vehicle (i.e., 8) & 9 sec & 8 sec & 11 sec \\
        \hline
        First class-1 vehicle (i.e., 1) & 0.5 sec & 0.5 sec & 0.5 sec \\
        \hline
        First class-2 vehicle (i.e., 4) & 4 sec & 5 sec & 2 sec \\
        \hline
    \end{tabular}
    \caption{System times for sequence in Fig.~\ref{fig:sequence}.}
    \label{tab:discharge}
\end{table}
One can interpret the comparison between MS and LQF as the trade-off between system-wide efficiency and fairness. MS may hold some vehicles for a long time and result in large variance in delay. LQF, however, tends to evenly distribute delay over vehicles. For example, vehicle 1, which is in the front of the class-1 queue, has to wait for 0 sec under all policies, while vehicle 4, which is in the front of the class-2 queue, has to wait for 4 sec, 5 sec, and 2 sec under FIFO, MS, and LQF, respectively.

%%%%%%%%%%%%%%%%%%%%%%%%%%%%%%%%%%%%%%%%%%%%%%%%%%
\subsection{Proof of Theorem~\ref{thm_two}}
\label{sub_proof}

% We construct policy-specific Lyapunov functions that verify the comparison theorem for Markov processes \cite[Theorem 4.3]{meyn1993stability}.

\subsubsection{FIFO}
\label{sub_fifo}
This proof consists of three parts: (i) sufficiency of \eqref{eq_fifo} and the upper bound $\overline{W}_1$, (ii) necessity of \eqref{eq_fifo}, and (iii) the lower bound $\overline{W}_0$.

\begin{figure}[hbt]
    \centering
    \includegraphics[width=0.25\textwidth]{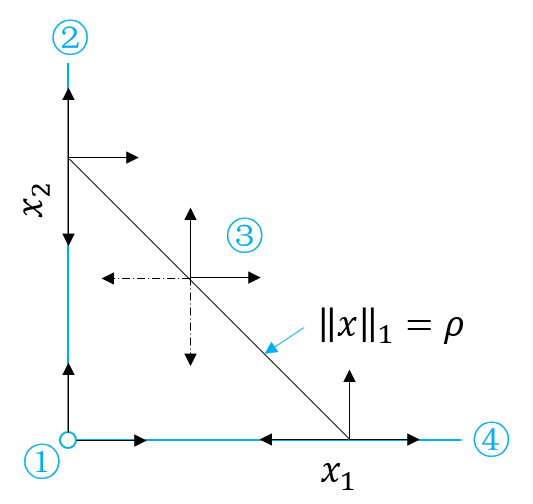}
    \caption{Regimes 1--4 for verifying drift condition under FIFO.}
    \label{1norm}
\end{figure}

To show {\bf sufficiency} of \eqref{eq_fifo} and the {\bf upper bound} $\overline{W}_1$, consider the Lyapunov function
\begin{align}
    V_1(g,u)=\frac12\|x\|_1^2+a_y\|x\|_1,
    \label{eq_V1}
\end{align}
where
\begin{align*}
    a_y=\frac{\lambda_1(\theta_{y,1}-\theta_{-y,1})+\lambda_2(\theta_{y,2}-\theta_{-y,2})}{2(\lambda_1+\lambda_2)},
    \quad y\in\mathcal K
\end{align*}
are in fact distinct elements in the matrix $A(\lambda)$; note that $x$ and $y$ can be derived from $(g,u)$.
% We apply the generator $\mathscr A_1$ to the Lyapunov function to compute the mean drift.
The main challenge for the proof is to show that the mean drift satisfies
\begin{align}
    \mathscr A_1V_1(g,u)\le-c_1\|x\|_1+d_1,
    \quad\forall (g,u)\in\mathcal G\times \mathcal U,
    \label{eq_drift1}
\end{align}
for constants $c_1>0$ and $d_1<\infty$; note that the generator also depends on the policy.
% The drift condition \eqref{eq_drift1} is essential to apply the comparison theorem for Markov processes \cite[Theorem 4.3]{meyn1993stability}, which leads to sufficiency of \eqref{eq_fifo} and the upper bound.
We verify \eqref{eq_drift1} over the four qualitatively different regimes in Fig.~\ref{1norm}.
Regime 1 is the singleton $\{0\}$; based on properties of PDMPs \cite{benaim15,cloez15}, the mean drift is given by
{\begin{align}
    &\mathscr A_1V_1=\sum_{k=1}^2\lambda_k\mathrm E\Big[\frac12(\theta_{y,k}+R)^2+a_y(\theta_{y,k}+R)\Big]
    \nonumber\\
    &=\sum_{k=1}^2\lambda_k\Big(\frac12\theta_{y,k}^2+\theta_{y,k}\overline{R}+\frac12(\overline{R}^2+\sigma_R^2)+a_y(\theta_{y,k}+\overline{R})\Big),\nonumber
\end{align}}%
which leads to
\begin{align}
    \mathscr A_1V_1\le\Big\|&\Big((\Theta+\overline RJ_{2,2})\circ(\frac12(\Theta+\overline RJ_{2,2})+A)\nonumber\\
    &+\frac12\sigma_R^2J_{1,2}\Big)\lambda\Big\|_\infty;
    \label{eq_d1}
\end{align}
In fact, the above holds over all four regimes.
Regimes 2--4 are given by $\{x:x_1=0,x_2>0\}$, $\{x:x_1>0,x_2>0\}$, $\{x:x_1>0,x_2=0\}$, respectively. The main difference between these regimes is that $y$ must be 2 (resp. 1) over regime 2 (resp. 4), while $y$ can be either 1 or 2 over regime 3. The mean drift over regimes 2--4 satisfies
\begin{align}
    &\mathscr A_1V_1=\Big(-1+\sum_{k=1}^2\lambda_k(\theta_{y,k}+\bar R)+\lambda_{-y}(a_{-y}
    \nonumber\\
    &-a_y)\Big)\|x\|_1+\frac12\sum_{k=1}^2\lambda_k\mathrm E[(\theta_{y,k}+R)^2]-a_y.
    \label{eq_AmV1}
\end{align}
Note that $\theta_{y,k}$ in the summations means that when a class-$k$ vehicle arrives, it has to keep a headway of $\theta_{y,k}$ away from the last vehicle, which is of class $y$.
Substitution of $a_y$ into \eqref{eq_AmV1} yields
\begin{align}
    &-1+\sum_{k=1}^2\lambda_k(\theta_{y,k}+\overline R)+\lambda_{-y}(a_{-y}-a_y)
    \nonumber\\
    &=-1+\sum_{\ell=1}^2\frac{\lambda_\ell}{\|\lambda\|_1}\sum_{k=1}^2\lambda_k(\theta_{\ell,k}+\overline R),
    \quad y=1,2.
    \label{eq_-1+sum}
\end{align}
Then, one can obtain from \eqref{eq_d1} and \eqref{eq_-1+sum} that there exist
\begin{align*}
    c_1&:=1-\Big(\frac{\lambda^T}{\|\lambda\|_1}\Theta+\overline RJ_{1,2}\Big)\lambda>0,\\
    d_1&:=\frac12\Big(\mathrm{diag}\Big((\Theta+\overline RJ_{2,2})\circ(\Theta+\overline RJ_{2,2})\Big)+\sigma_R^2J_{1,2}\Big)\lambda\\&<\infty
\end{align*}
such that \eqref{eq_drift1} holds.
Then, we can apply the comparison theorem \cite[Theorem 4.3]{meyn1993stability} to obtain boundedness in the sense of \eqref{eq_zeta} with $\overline{W}=d_1/c_1$ and the upper bound in \eqref{eq_Wbar1}.

Note that the limit $\overline X$ in \eqref{eq_Xbar} exists under most practical policies if $X(t)$ is bounded in the sense of \eqref{eq_zeta}.
The key is to argue that the PDMP is irreducible in the sense of \cite[p.5]{meyn1993survey}.
The irreducibility argument in the discrete state space is straightforward. For the continuous state space, note that, under most practical policies, the state $x=[0,0]$ (or $u=\{0,0\}$) can be exactly attained from any initial condition in finite time with a positive probability. Hence, the PDMP is irreducible. Since the PDMP is also bounded in the sense of \eqref{eq_zeta}, the PDMP is positive Harris \cite[Theorem 7]{meyn1993survey}, and thus the limit in \eqref{eq_Xbar} exists.

To show {\bf necessity} of \eqref{eq_fifo}, we a consider bounded test function
\begin{align*}
    W_1(g,u)=1- e^{-\alpha_1(\|x\|_1+\beta_y)},
\end{align*}
where $\alpha_1,\beta_y$ are positive constants. In the following, we show that 
\begin{align}
    \mathscr A_1W_1\ge0,\quad\forall (g,u)\in\mathcal G\times\mathcal U.
    \label{eq_AWi}
\end{align}
If $x\ne 0$, the mean drift is given by
\begin{align*}
    \mathscr A_1W_1=&- e^{-\alpha_1(\|x\|_1+\beta_y)}(-1)+\sum_{k=1}^2\lambda_k(e^{-(\alpha_1\|x\|_1+\beta_y)}\\
    &-\mathrm E[e^{-\alpha_1(\|x\|_1+\theta_{y,k}+R+\beta_k)}])\\
    =&e^{-\alpha_1(\|x\|_1+\beta_y)}\Big(-1+\sum_{k=1}^2\lambda_k(1\\
    &-e^{-\alpha_1\theta_{y,k}}g_R(-\alpha)e^{-\alpha_1(\beta_k-\beta_y)}\Big)
\end{align*}
where $g_R(-\alpha_1)$ is the MGF of the rv $R$.
We can expand the exponential terms as follows:
\begin{align*}
    &1-e^{-\alpha_1\theta_{y,k}}g_R(-\alpha_1)e^{-\alpha_1(\beta_k-\beta_y)}\\
    &=
    \alpha_1(\theta_{y,k}+\overline R+\beta_k-\beta_y)+o(\alpha_1),
\end{align*}
where $\beta_y$ are balancing terms analogous to $a_y$ in \eqref{eq_V1}.
Hence, if 
$
        ({\lambda^T}/{\|\lambda\|_1}\Theta+\bar RJ_{1,2})\lambda>1
$
then there exists $\alpha>0$ such that
$$
1-e^{-\alpha_1\theta_{y,k}}g_R(-\alpha_1)e^{-\alpha_1(\beta_k-\beta_y)}\ge0,
$$
which leads to \eqref{eq_AWi}.
Thus, we can apply the drift criterion for transience \cite[Theorem 4]{meyn1993survey} and conclude that the PDMP is unstable. The case that 
$
       ({\lambda^T}/{\|\lambda\|_1}\Theta+\bar RJ_{1,2})\lambda=1
$
involves null-recurrence type arguments, which we omit here; in fact, this boundary case is of limited practical relevance, since an arbitrarily small deviation of model parameters will prevent this from happening.

The {\bf lower bound} $\overline W_0$ results from the construction of an M/G/1 process that optimistically estimates the traffic discharging process. The M/G/1 process has an arrival rate of $\|\lambda\|_1$. The service time is the sum of two independent rv s $\Phi$ and $R$, where $\Phi$ has the probability mass function
\begin{align*}
    p_\Phi(\theta_{11})=\frac{\lambda_1}{\|\lambda\|_1},\quad
    p_\Phi(\theta_{22})=\frac{\lambda_2}{\|\lambda\|_1}
\end{align*}
and $R$ has the CDF $F_R(r)$.
This M/G/1 process essentially uses $\theta_{k,k}$ to under-estimate the headway before a class-$k$ vehicle, which thus leads to an under-estimate of delay. The expression for $\overline W_0$ results from Pollaczek-Khinchin formula \cite[p.248]{gallager2013stochastic}, and this lower bound applies to any sequencing policy.

%%%%%%%%%%%%%%%%%%%%%%%%%%%%%%%%%%%%%%%%
\subsubsection{MS}
To show {\bf sufficiency} of \eqref{eq_ms} and the {\bf upper bound} $\overline{W}_2$, we consider a ``decomposed'' process $(\tilde X,F(t))\in\mathbb R_{\ge0}^2\times\mathbb R_{\ge0}$ such that $\|\tilde X(t)\|_1+F(t)=\|X(t)\|_1$ for all $\ge0$, where $\tilde X(t)$ tracks the ``crossing time-to-go'' and $F(t)$ tracks the ``switchover time-to-go''. Therefore, the decomposed process is stable if and only if the original process is stable.
Specifically, the decomposed process initiates according to
$$
\tilde X(0)=X(0),
\ 
F(t)=0.
$$
Then, for $t>0$, if $\tilde X_k(t)>0$ and if $F(t)=0$, a class-$k$ arrival increases $X_k(t)$ by $\theta_{k,k}+R$, and $F(t)$ is unchanged. Let $\epsilon$ be a constant such that $0<\epsilon<1-\mathrm{diag}(\Theta+\overline{R}J_{2\times2})\lambda$; note that \eqref{eq_ms} ensures the existence of $\epsilon$. If $\tilde X_k(t)=0$, then a class-$k$ arrival increases $\tilde X_k(t)$ by $\theta_{k,k}+\epsilon+R$, and $F(t)$ is increased by $(\theta_{-k,k}-\theta_{k,k}-\epsilon)_+$. Whenever $F(t)>0$, a class-$k$ arrival will increase $\tilde X(t)$ by $\theta_{k,k}+\epsilon+R$ and decrease $F(t)$ by $\min\{F(t),\epsilon\}$.
By definition, $\tilde X(t)\in\mathbb R_{\ge0}^2$ and $F(t)\in[0,\bar f]$, where $\bar f=\sum_k(\theta_{-k,k}-\theta_{k,k})$.
Thus, $\tilde X(t)$ is bounded if and only if $X(t)$ is bounded.
Furthermore, $\{Z(t),\tilde X(t),F(t);t>0\}$ itself is a Markov process.
Then, consider the Lyapunov function
\begin{align}
    V_2(z,\tilde x,f)=\frac12\|\tilde x\|_1^2.
    \label{eq_V2}
\end{align}
For any $(z,\tilde x,f)\in\mathcal K\times\mathbb R_{\ge0}^2\times[0,\bar f]$, the mean drift is
\begin{align}
    &\mathscr A_2V_2(z,\tilde x,f)=\Big(-1+\sum_{k=1}^2\lambda_k(\theta_{k,k}+\bar R+\epsilon\mathbb I\{f\nonumber\\
    &>0\})\Big)\|x\|_1+\frac12\sum_{k=1}^2\lambda_k\mathrm E[(\theta_{k,k}+R+\epsilon\mathbb I\{f>0\})^2].
    \label{eq_AmV2}
\end{align}
The use of $\epsilon$ (and thus the decomposed process) is to distribute the switchover-induced increment $(\theta_{k,-k}-\theta_{k,k})$ to the $\ceil{(\theta_{k,-k}-\theta_{k,k})/\epsilon}$ subsequent arrivals, so that the mean drift is still negative upon switchovers.
% The main differences between the above and \eqref{eq_AmV1} include: (i) the headway is $\theta_{k,k}$ in the above, since all incoming vehicles will be placed after a vehicle of the same class; (ii) the balancing terms $a_y$ do not exist.
Then, by analogy with the FIFO case, we can show that there exist
\begin{align*}
    c_2:=&1-\mathrm{diag}(\Theta+\overline RJ_{2,2})\lambda-\epsilon>0,\\
    d_2:=&\frac12\Big(\mathrm{diag}\Big((\Theta+(\overline R+\epsilon)J_{2,2})\circ(\Theta+(\overline R+\epsilon)J_{2,2})\Big)\nonumber\\
    &+\sigma_R^2J_{1,2}\Big)\lambda<\infty
\end{align*}
such that for all $(z,\tilde x,f)\in\mathcal K\times\mathbb R_{\ge0}^2\times[0,\bar f]$,
\begin{align*}
    \mathscr A_2V_2(z,\tilde x,f)\le-c_2\|\tilde x\|_1+d_2.
\end{align*}
% Note that if \eqref{eq_ms} holds, then there exists $\epsilon>0$ such that $c_2>0$.
Also note that we can let $\epsilon\downarrow0$, since there is at most one switchover to go under MS; this argument would be invalid under other policies (e.g., FIFO), since there could be infinitely many switchovers to go. Finally, $\|\tilde X(t)\|_1$ is associated with the upper bound $d_2/c_2$, and thus $\|X(t)\|$ is associated with the upper bound given by $\overline W_2$; the difference between $\overline W_2$ and $d_2/c_2$ is the mean switchover time under FIFO, which is indeed an upper bound for that under MS.

To show {\bf necessity} of \eqref{eq_ms}, consider the test function 
\begin{align*}
    W_2(g,u)=1-e^{\alpha_2\|x\|_1}.
\end{align*}
Then, one can analogously to the FIFO case.

%%%%%%%%%%%%%%%%%%%%%%%%%%%%%%%%%%%%%%%%
\subsubsection{LQF}
\begin{figure}[hbt]
    \centering
    \includegraphics[width=0.3\textwidth]{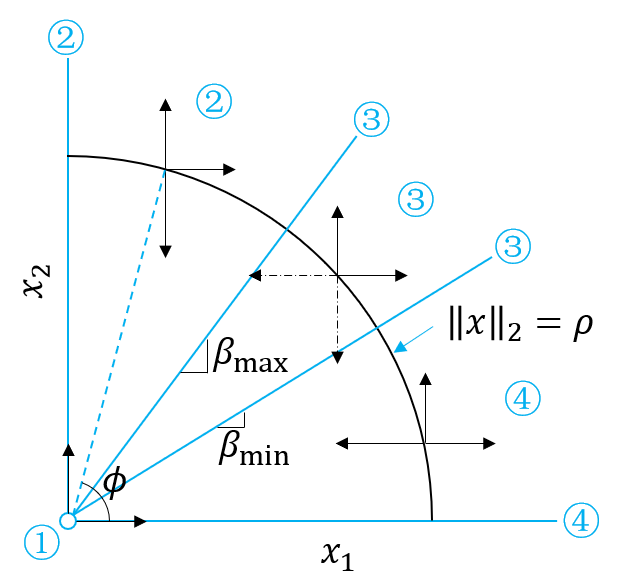}
    \caption{Regimes 1--4 for verifying drift condition under LQF.}
    \label{2norm}
\end{figure}

Consider the Lyapunov function
\begin{align*}
    V_{3}(g,u)=\frac12\|x\|_2^2.
\end{align*}
To show that $V_{3}$ drifts negatively, we essentially need to show that the ``mean velocity'' of the PDMP points towards the interior of the level curves; see Fig.~\ref{2norm}.
The drift condition involves four qualitatively different regimes, which are labeled 1--4 in Fig.~\ref{2norm} with
\begin{align*}
    \beta_{\min}=\frac{\lambda_1\tilde\theta_{21}}{1-\lambda_2\tilde\theta_{22}},\quad
    \beta_{\max}=\frac{1-\lambda_1\tilde\theta_{11}}{\lambda_2\tilde\theta_{12}};
\end{align*}
note that $0<\beta_{\min}<\beta_{\max}<\infty$ if \eqref{eq_lqf} holds.
Also note that each regime for $x$ corresponds to a unique regime for $(g,u)$.
Next, we consider the four regimes separately.

Regime \textcircled{1}: $x_1=x_2=0$: By analogy with FIFO, we have
    \begin{align}
        \mathscr A_3V_3&\le
        \Big\|\frac12\Big((\Theta+\overline RJ_{2,2})\circ(\Theta+\overline RJ_{2,2})\nonumber\\
        &+\sigma_R^2J_{1,2}\Big)\lambda\Big\|_\infty=:d_3<\infty
    \end{align}
    for all regimes and thus indeed for regime 1.
    
Regime \textcircled{2}: $x_1\ge0$, $x_2>\beta_{\max} x_1$: 
    We require the matrix $B(\lambda)$ as defined in \eqref{eq_B} to characterize the mean drift to account for two peculiarities of the LQF policy:
    \begin{enumerate}
        \item Since class 2 has a longer queue, a new class-2 arrival will be placed at the end of the sequence of crossing, which leads to a service time of $\theta_{22}+R$. A class-1 arrival will be placed either after a class-1 vehicle or after a class-2 vehicle. In the former sub-case, the arrival has a service time of $\theta_{11}+R$. In the latter sub-case, the arrival has a service time of $\theta_{21}+R$. In addition, the arrival may also be inserted in front of a class-2 vehicle, which will further increases the service time of the following class-1 vehicle from $\theta_{22}+R$ to $\theta_{12}+R$; see Fig.~\ref{fig:two_lqf}.
        \begin{figure}[hbt]
    \centering
    \includegraphics[width=0.45\textwidth]{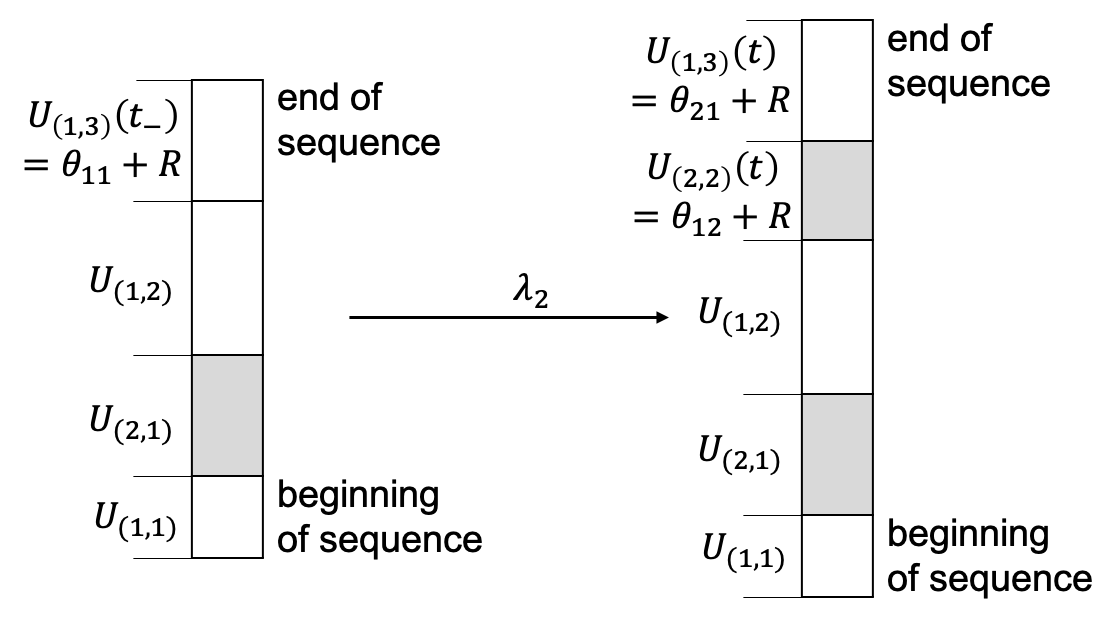}
    \caption{Under LQF, a new arrival (2,2) may affect the service time of an existing vehicle (1,3).}
    \label{fig:two_lqf}
    \end{figure}
        
        \item Although the LQF policy essentially prioritizes the longer queue for discharging (i.e., class 2 in thisregime), either class 1 or 2 can be discharged in this regime. The reason is that an ongoing crossing cannot be interrupted; the state may switch between regimes as a crossing goes on. To resolve this, we assume that, in this regime, a class-1 arrival leads to an additional ``phantom'' increment of $u_1$ in $x_1$; this increment will decrease and vanish synchronously with the ongoing crossing, so it will also contributes negatively to the mean drift (see Fig.~\ref{2norm}). Since $u_1\le\theta_{21}+R_{\max}$, such increment is also upper-bounded by $\theta_{21}+R_{\max}$.
        
    \end{enumerate}
To sum up, a class-1 arrival will increase $x_1$ by no greater than $\theta_{21}+R+\theta_{12}+R_{\max}$ and increase $x_2$ by no greater than $\theta_{12}-\theta_{22}$; a class-2 arrival will not influence $x_1$ and will increase $x_2$ by $\theta_{22}+R$.
Thus, the mean drift is upper-bounded by
    \begin{align*}
        &\mathscr A_3V_3\le\bigg(\lambda_1(\theta_{21}+R+\theta_{12}+R_{\max})\cos\phi+\Big(\lambda_1(\theta_{12}
        \\
        &\quad-\theta_{22})+\lambda_2(\theta_{22}+\overline{R})-1\Big)\sin\phi+\frac{o(\| x\|_2)}{\| x\|_2}\bigg)\| x\|_2,
    \end{align*}
    where $\phi=\arctan(x_2/x_1)$; see Fig.~\ref{2norm}. Note that
    \begin{align*}
        &\lambda_1(\theta_{21}+R+\theta_{12}+R_{\max})\cos\phi+\Big(\lambda_1(\theta_{12}-\theta_{22})
        \\
        &\quad+\lambda_2(\theta_{22}+\overline{R})-1\Big)\sin\phi
        \\
        &\stackrel{\eqref{eq_lqf}}{\le}\lambda_1(\theta_{21}+R+\theta_{12}+R_{\max})\frac{1}{\sqrt{1+\beta_{\max}^2}}
        \\
        &\quad+\lambda_2(\theta_{22}+\overline{R})-1\Big)\frac{\beta_{\max}}{\sqrt{1+\beta_{\max}^2}}\\
        &\stackrel{\eqref{eq_beta}}{\le}\frac{\mathrm{det}(\tilde\Theta+\overline{R}J_{2,2})\lambda_1\lambda_2+\mathrm{diag}(\tilde\Theta)\lambda-1}{\sqrt{(1-\lambda_1\tilde\theta_{11})^2+(\lambda_2\theta_{12})^2}}\\
        &=:-c_{3,1}<0,
    \end{align*}
    where also defines the constant $c_{3,1}$.
     Thus, if $\|x\|_2$ is sufficiently large, we have
     \begin{align*}
         \mathscr A_3V_3\le-(c_{3,1}-\epsilon)\|x\|_2,
     \end{align*}
     where $\epsilon>0$ is an arbitrarily small number. Hence, given \eqref{eq_lqf}, we have
     \begin{align*}
         \mathscr A_3V_3<-c_{3,1}\|x\|_2+d_3
     \end{align*}
     over regime 2.
    
Regime \textcircled{3}: $x_1>0$, $\beta_{\min} x_1\le x_2\le\beta_{\max} x_1$: Either class is discharged.
    If class 2 is discharged, the proof is analogous to regime 2. If class 1 is discharged, we have
    \begin{align*}
        \mathscr A_3V_3=&\bigg(\lambda_1(\theta_{11}+\overline{R}-1)\cos\phi+\lambda_2\Big((\theta_{12}+\theta_{21}\\
        &-\theta_{11})+\overline{R}\Big)\sin\phi+\frac{o(\| x\|_2)}{\| x\|_2}\bigg)\| x\|_2
        \\
        \le&\bigg(\lambda_1(\theta_{11}+\overline{R}-1)\frac{1}{\sqrt{1+\beta_{\min}^2}}+\lambda_2\Big((\theta_{12}+\theta_{21}\\
        &-\theta_{11})+\overline{R}\Big)\frac{\beta_{\min}}{\sqrt{1+\beta_{\min}^2}}+\frac{o(\| x\|_2)}{\| x\|_2}\bigg)\| x\|_2\\
        \stackrel{\eqref{eq_beta}}\le&\Bigg(\frac{\mathrm{det}(\tilde\Theta+\overline{R}J_{2,2})\lambda_1\lambda_2+\mathrm{diag}(\tilde\Theta)\lambda-1}{\sqrt{(1-\lambda_1\tilde\theta_{11})^2+(\lambda_2\theta_{12})^2}}\\
        &+\frac{o(\| x\|_2)}{\| x\|_2}\Bigg)\|x\|_2\\
        =&-\Bigg(c_{3,2}-\frac{o(\| x\|_2)}{\| x\|_2}\Bigg)\|x\|_2<0,
    \end{align*}
    which also defines the constant $c_{3,2}$.
    Thus, if $\|x\|_2$ is sufficiently large, we have
     \begin{align*}
         \mathscr A_3V_3\le-(c_{3,2}-\epsilon)\|x\|_2,
     \end{align*}
     where $\epsilon>0$ is an arbitrarily small number. Hence, given \eqref{eq_lqf},
     \begin{align*}
         \mathscr A_3V_3<-c_{3}\|x\|_2+d_3
     \end{align*}
     over regime 2, where $c_3=\min\{c_{3,1},c_{3,2}\}$.
     
Regime \textcircled{4}: $x_2\ge0$, $x_1>\beta_{\min}x_1$: analogous to regime 2.

In conclusion, given \eqref{eq_lqf}, we have
\begin{align*}
    \mathscr A_3V_3\le-c_3\|x\|_2+d_3
\end{align*}
over all regimes. By the comparison theorem, we have
\begin{align*}
    d_3/c_3\ge \|x\|_2\ge\|x\|_1/\sqrt2,
\end{align*}
which leads to stability as well as the upper bound $\overline{W}_3$.

%% file: sections/35_validation.tex
\section{Implementation and Validation}
\label{sec_simulation}

In this section, we discuss the implementation of the PDMP-based sequencing policies in practice and validate the theoretical results via simulation. Section \ref{sub:impl} introduces how we translate the PDMP-based control to the sequence and set time windows for incoming vehicles to cross the intersection. Section \ref{sub:two_sim} shows how to implement the set times of arrival in SUMO and discuss the simulation results. Table \ref{table:sim}  lists the variables involved in the implementation of the sequence policies.

\begin{table}[hbt]
\centering
\begin{tabular}{cl}
\hline
Notation & Variable [unit]\\
\hline
$\lambda_k$ & arrival rate [veh/sec]\\
\hline
$\theta_{ij}$ & minimal headway [sec]\\
\hline
$\Theta $& headway matrix [sec]\\
% \hline
% \makecell[c]{$N_k(t)$\\}
% & \makecell[c]{the number of class-$k$ vehicles waiting \\ for discharge  at time $t$ / class-$k$ count [veh]}\\
\hline
$N(t)$ & total count [veh]\\
\hline
$\delta$ & simulation time step size [sec]\\
\hline
$G(t)$ & sequence of crossing [($k_i,n_i$)]\\
\hline
$T^{i}_{\mathrm{ms}}$ &minimal set time for the $i$th vehicle to cross [sec] \\
\hline
$T^{i}_{\mathrm{min}}$ & minimal time for the $i$th vehicle to cross [sec]\\
\hline
$T^{i}_{\mathrm{set}}$ & set time for the $i$th vehicle to cross [sec]\\
\hline
$T^{i}_{\mathrm{e}}$ & time of the $i$th vehicle arrival [sec]\\
\hline
$t$ & current time in simulation [sec]\\
\hline
$L$ & length of approaching zone [m]\\
\hline
$v_{\mathrm{max}}$ & nominal/maximum speed [m/sec]\\
\hline
$v^i$ & speed for the $i$th vehicle [m/sec]\\
\hline
$a_+$ &  acceleration [m/se$\mbox{c}^2$]\\
\hline
$a_-$ & deceleration  [m/se$\mbox{c}^2$]\\
\hline
\end{tabular}
\caption{Notations used in simulation.}
\label{table:sim} 
\end{table}

%%%%%%%%%%%%%%%%%%%%%%%%%%%%%%%%%%%%%%%%%%%%%%%%%
\subsection{Implementation of PDMP-based policies }
\label{sub:impl}
We control the vehicles by setting their times to cross; that is, all vehicles will cross the intersection at their set times. We assume that all vehicles arrive with the same speed $v_{\mathrm{max}}$. The minimal set time for the $i$th vehicle (of all vehicles) to cross can be calculated by  
\begin{equation*}
    T_{\mathrm{ms}}^{G_i}=\frac{L}{v_{\mathrm{max}}}+T_{\mathrm{e}}^{G_i}.
\end{equation*}
Note that $G_i=(k_i,n_i)$, i.e. the $n$th class $k$ vehicle. The set time for each vehicle to cross is decided by the sequence of crossing $G(t)$
\begin{subequations}
    \begin{align}
        T_{\mathrm{set}}^{G_1}&=T_{\mathrm{ms}}^{G_1}, \label{eq:tset1} \\
        T_{\mathrm{set}}^{G_i}&=max(T_{\mathrm{ms}}^{G_i},T_{\mathrm{set}}^{G_{i-1}}+\theta_{k_ik_{i-1}}+\overline{R}) \quad i=2,3... \label{eq:tset2}
    \end{align}
 \end{subequations}
\subsubsection{FIFO}
For the FIFO policy, the sequence of vehicles to cross is the sequence of arrival. Once the sequence of crossing is decided, we can easily calculate the set times by \eqref{eq:tset1}-\eqref{eq:tset2}. Algorithm $\ref{algo:FIFO}$ is the pseudo-code to calculate the crossing time windows for vehicles.

\begin{algorithm}[hbt]
	\caption{Crossing time windows of FIFO} 
        \label{algo:FIFO}
	\LinesNumbered  
	\KwIn{ \emph{$\Theta$}, \emph{$\overline{R}$}}
	\KwOut{\emph{G}, \emph{$t$}, \emph{$T_{\mathrm{set}}$}} 
        $t \leftarrow 0$\\
        $G \leftarrow $  empty array\\
        $T_{\mathrm{ms}} \leftarrow $ empty array\\
        \While {Simulation is running}{
        $t \leftarrow t+1$\\
        \If{new vehicle $(k_i,n_i)$ arrives}{
           $ G_i\leftarrow (k_i,n_i)$ \\
            \eIf{$i=0$}{
            $T_{\mathrm{set}}^{G_i}\leftarrow T_{\mathrm{ms}}^{G_i}$ \\
            }{
            $T_{\mathrm{set}}^{G_i} \leftarrow max(T_{\mathrm{ms}}^{G_i},T_{\mathrm{set}}^{G_{i-1}}+\theta_{k_ik_{i-1}}+\overline{R})$ \\
            }
       
        }
         \textbf{Return} $G$, $t$, $T_{\mathrm{set}}$
        }

\end{algorithm}

\subsubsection{MS}
For the MS policy, when a new vehicle (labeled as $(k_i,n_i)$) arrives, we need to first find $(k_i,n_{i-1})$.  We assume $G_f=(k_i,n_{i-1})$; that is, $(k_i,n_{i-1})$ is the $f$th vehicle in the sequence of crossing.
\begin{enumerate}
    \item If $T_{\mathrm{ms}}^{G_f} \leq T_{\mathrm{set}}^{G_f}+\theta_{11}+\overline{R}$, then we shift every vehicle following $G_f$ to the next position in the sequence, and let $G_{f+1}=(k_i,n_i)$.
    \item If $T_{\mathrm{ms}}^{G_f} > T_{\mathrm{set}}^{G_f}+\theta_{11}+\overline{R}$, then let $j=f$. We check if $T_{\mathrm{ms}}^{G_f} \leq T_{\mathrm{set}}^{G_{j+1}]} +\theta_{12}+\overline{R}$ and 2*$(\theta_{12}+\overline{R}) \leq T_{\mathrm{set}}^{G_{j+1}}-T_{\mathrm{set}}^{G_j} $. Then we shift every vehicle following $G_j$ to the next position in the sequence, and let $G_{j+1}=(k_i,n_i)$, or else we let $j=j+1$ and repeat this until the  vehicle is inserted into the queue.
\end{enumerate}
 Once the sequence to cross is decided, we can compute $T_{\mathrm{set}}$ using \eqref{eq:tset1}-\eqref{eq:tset2}. The algorithm is analogous to Algorithm $\ref{algo:FIFO}$.

\subsubsection{LQF}
For the LQF policy, each time there is a new vehicle enters or leaves the approaching zone, we will update $G$: we keep shifting vehicles from the longer class to $G$, until current releasing class is shorter than the other class and then we will repeat until all vehicles in the approaching zone is in $G$. In case of a tie, we maintain the class of discharge: this is preferable, since the number of switch-overs is reduced. Once the sequence to cross is decided, we can compute $T_{\mathrm{set}}$ using \eqref{eq:tset1}-\eqref{eq:tset2}. The algorithm is analogous to Algorithm $\ref{algo:FIFO}$.

%%%%%%%%%%%%%%%%%%%%%%%%%%%%%%%%%%%%%%%%%%%%%%%%%
\subsection{Simulation}
\label{sub:two_sim}

\begin{figure}[hbt]
  \centering
  \includegraphics[width=0.5\textwidth]{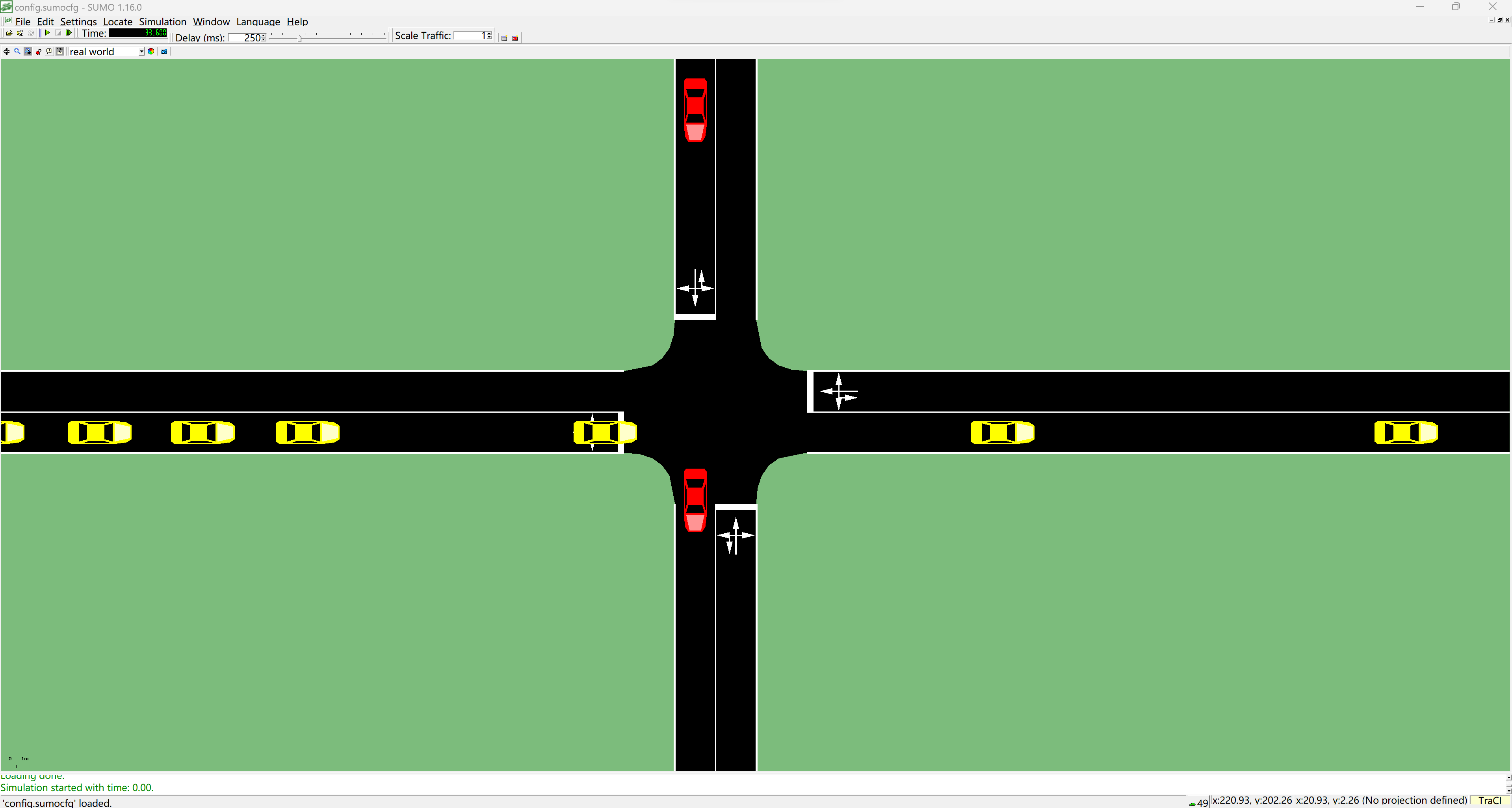}
  \caption{Simulation intersection in SUMO.}
  \label{fig:simulation_SUMO}
\end{figure}

We now validate the theoretical results by simulating the sequencing policies at the smart intersection. We applied various sequencing policies in Simulation of Urban Mobility (SUMO) \cite{krajzewicz2010traffic}; see
Fig. \ref{fig:simulation_SUMO}. 
The intersection consists of two directions, west-east (WE) and south-north (SN); each direction has an incoming vehicle flow. Vehicles are generated at random times with a minimal inter-arrival time of $\overline{R}$. The simulation step size is 0.1 sec, and a discrete-time Bernoulli process is used to approximate the continuous-time Poisson process considered in the theoretical analysis.

\subsubsection{Trajectory planning}
\begin{figure*}[htbp]
 \centering
 \subfigure[FIFO.]{
 \label{fig:subfig:a} 
 \includegraphics[width=2in]{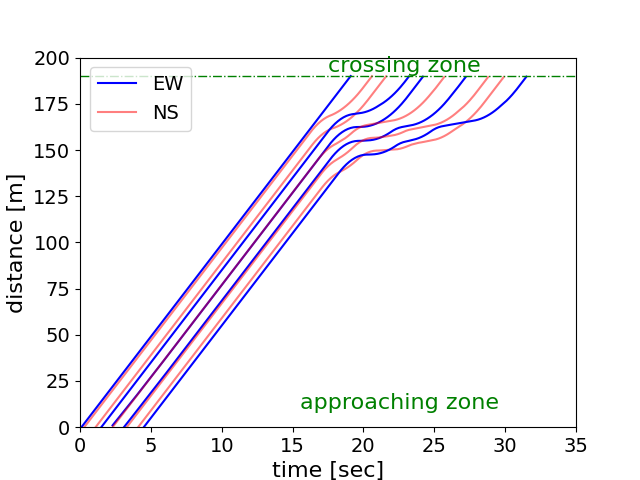}}
%  \hspace{0.01in}
 %
 \subfigure[MS.]{
 \label{fig:subfig:c} 
 \includegraphics[width=2in]{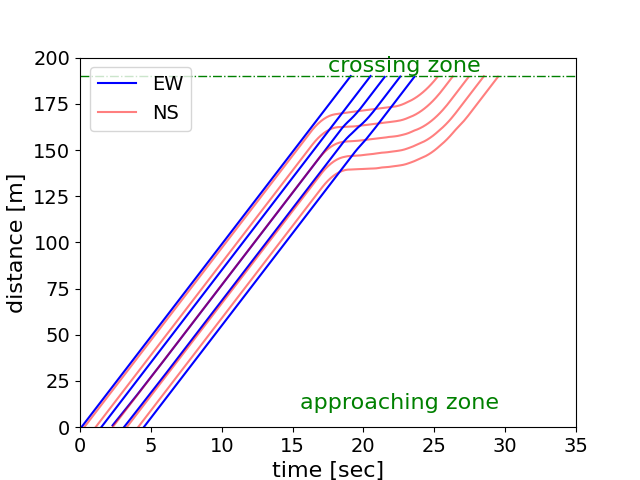}}
%  \hspace{0.01in}
 %
 \subfigure[LQF.]{
 \label{fig:subfig:b} 
 \includegraphics[width=2in]{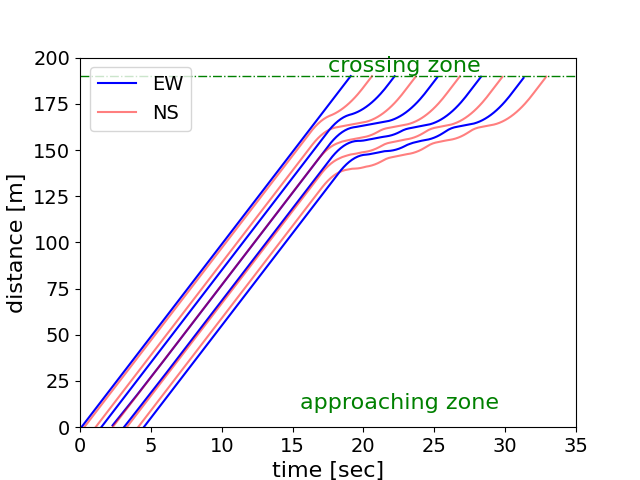}}
%  \hspace{0.01in}
 %
 \caption{Spatio-temporal trajectories of vehicles under various sequencing policies.}

 \label{fig:SUMO_distance_time} 
\end{figure*}

\begin{figure*}[htbp]
 \centering
 \subfigure[FIFO.]{
 \label{fig:subfig:a} 
 \includegraphics[width=1.9in]{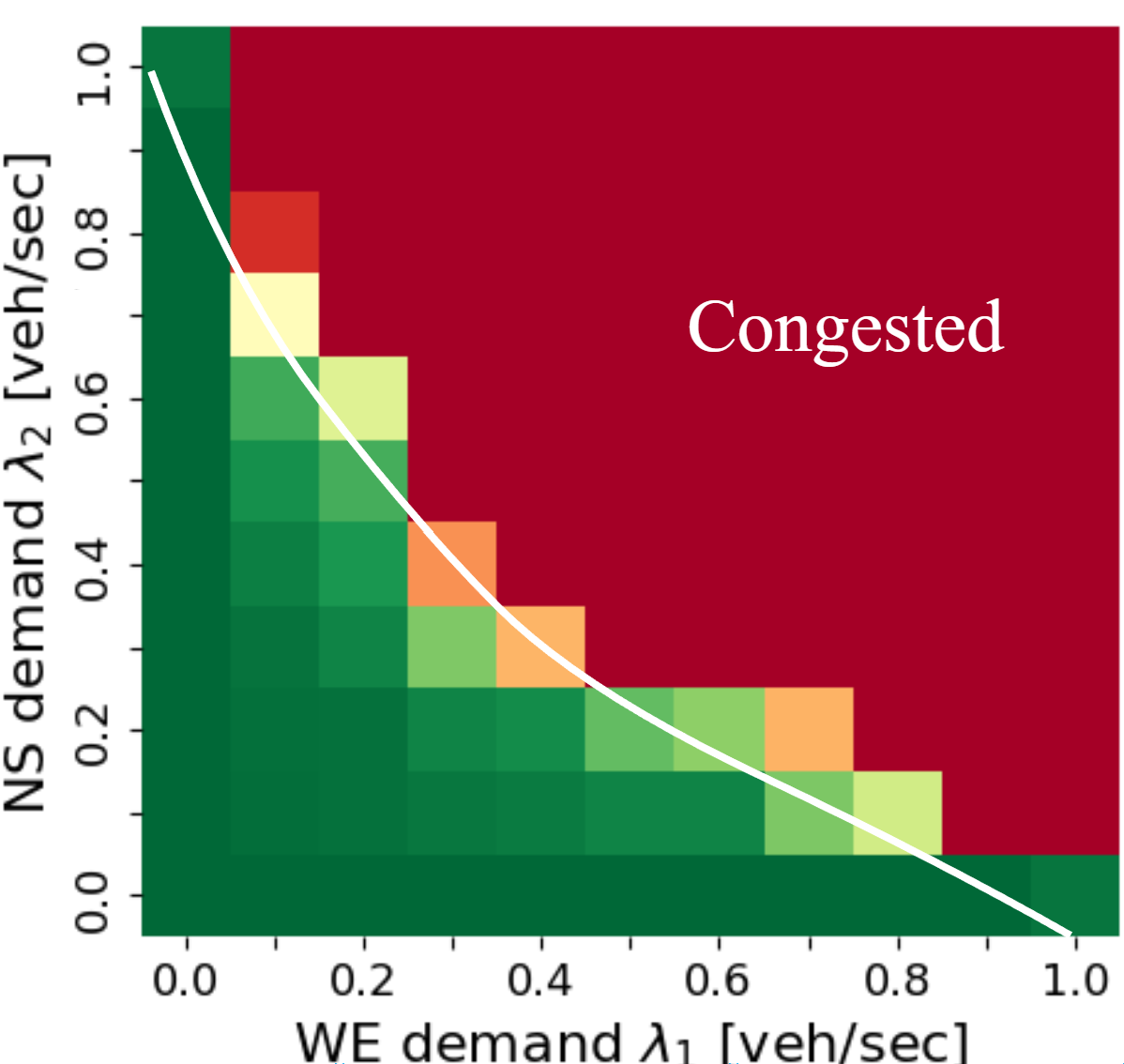}}
%  \hspace{0.01in}
 %
 \subfigure[MS.]{
 \label{fig:subfig:c} 
 \includegraphics[width=1.9in]{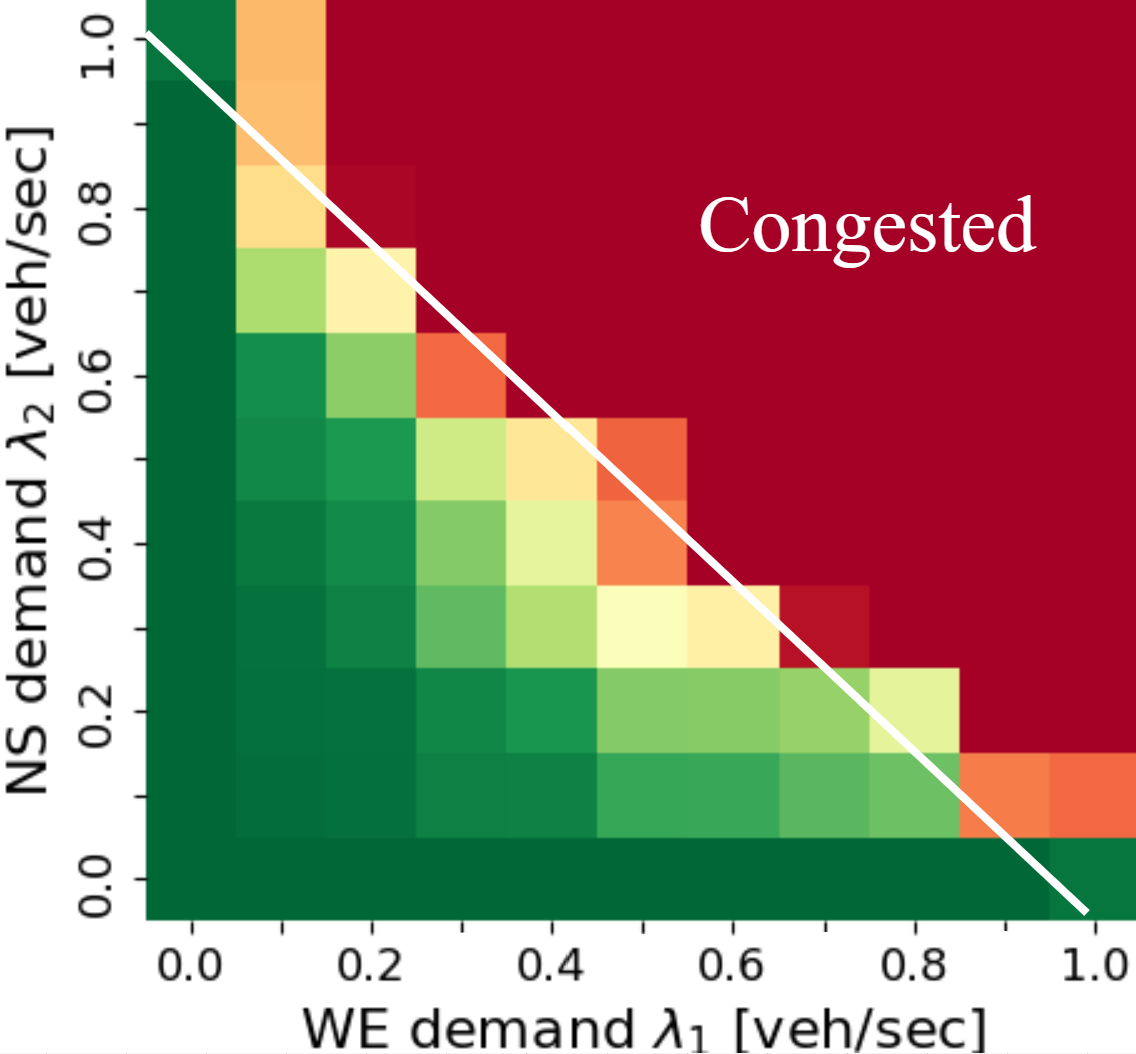}}
%  \hspace{0.01in}
 %
 \subfigure[LQF.]{
 \label{fig:subfig:b} 
 \includegraphics[width=2.3in]{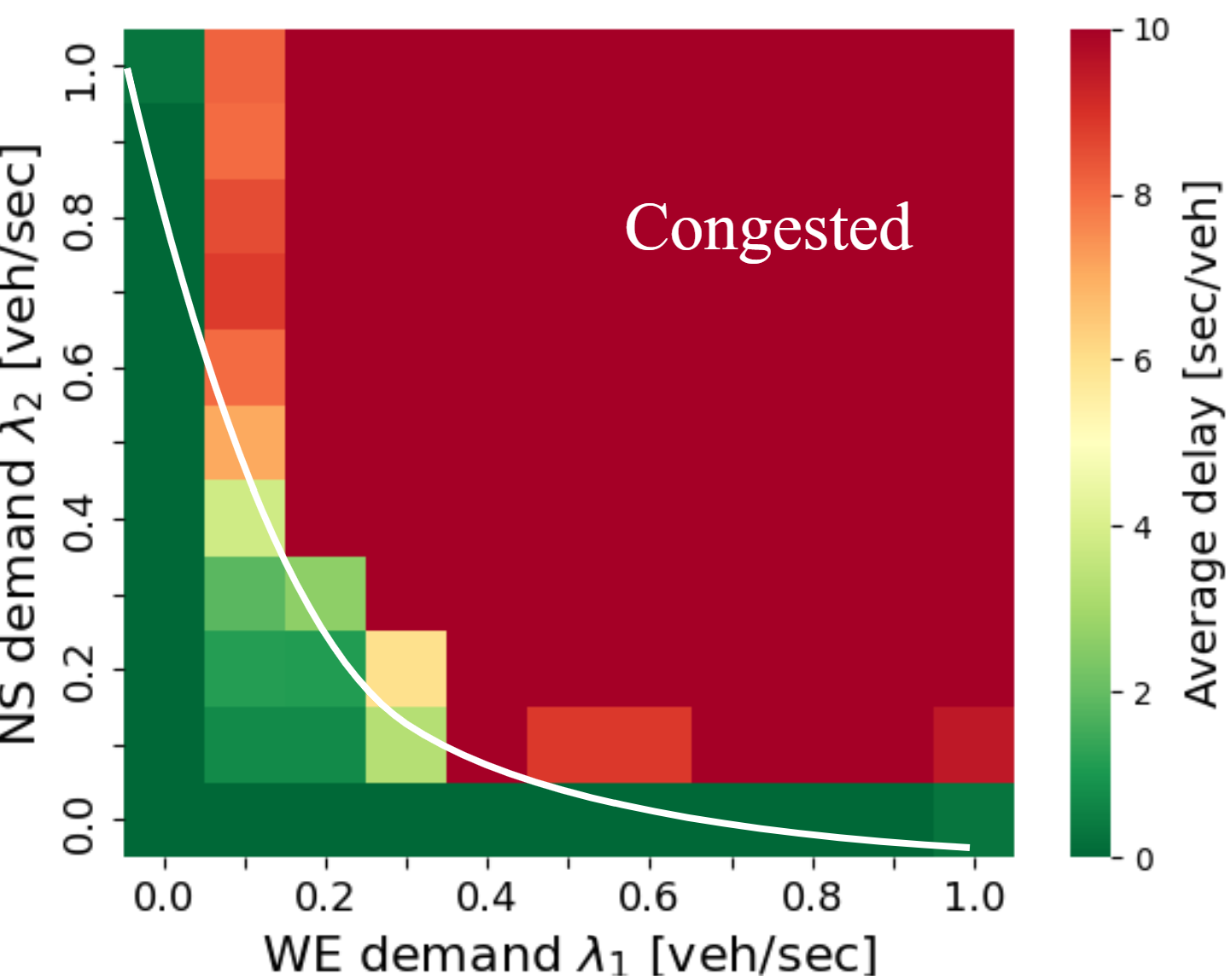}}
%  \hspace{0.01in}
 %

 \caption{Heat maps of average delay for different sequencing policies; the color bar applies to all three figures. The intersection is considered to be congested if the average delay attains 10 sec/veh. The white curves indicate the theoretical boundaries of the capacity regions.}

 \label{fig:SUMO_results} 
\end{figure*}

Using the sequence and set times obtained in the previous subsection, we consider a simple trajectory planning scheme such that the designated delay will be absorbed over the approaching zone. This scheme ensures higher crossing speed than holding vehicles at the intersection and is thus preferred: almost all vehicles will cross the intersection at the nominal speed, and the capacity is fully utilized.  After $G$ is decided, we calculate the set time to cross $T_{\mathrm{set}}$ and generate the time series of vehicle speed until crossing. For ease of presentation, we assume uniform acceleration/deceleration of vehicles; note that one can indeed replace such simple scheme with more sophisticated trajectory-planning schemes in the literature (e.g. \cite{ahn2017safety,2018A}). $T_{\mathrm{min}}^{i}$ is the minimal time to go (subject to safety constraints) for the $i$th vehicle to reach the intersection. $T_{\mathrm{min}}$ depends on whether the vehicle can attain the nominal speed before it reaches the intersection. $T_{\mathrm{min}}$ is calculated by the following formula:
\begin{equation*}
    T_{\mathrm{min}}^i=
    \begin{cases}
        \frac{(-v^i+\sqrt{(v^i)^2+2a_+d^i})}{a_+}+t \\
        \hspace{2.5cm} \mbox{if }\frac{v_{\mathrm{max}}^2-(v^i)^2}{2a_+}>d^i,\\
        \frac{(v_{\mathrm{max}}-v^i)}{a_+}+\frac{(d^i-\frac{v_{\mathrm{max}}^2-(v^i)^2}{2a_+})}{v_{\mathrm{max}}}+t \\ 
        \hspace{2.5cm} \mbox{if }\frac{v_{\mathrm{max}}^2-(v^i)^2}{2a_+}\leq d^i.
    \end{cases}
\end{equation*}
For the above formula, the first case is that the vehicle reaches the intersection when it is still accelerating. The second case is that the vehicle reaches the intersection with the speed $v_{\mathrm{max}}$.
The speed of the vehicle is bounded by $[0,v_{\mathrm{max}}]$, where $v_{\mathrm{max}}$ is the nominal speed. 
For each vehicle, if $T_{\mathrm{min}}^i<T_{\mathrm{set}}^i$, the vehicle needs to decelerate in order to avoid interference with the previous vehicle. The vehicle will decelerate at a constant acceleration $a_-$; if $T_{\mathrm{min}}^i>T_{\mathrm{set}}^i$, the vehicle cannot enter the crossing zone at the set time and we will let $T_{\mathrm{min}}^i=T_{\mathrm{set}}^i$; if $T_{\mathrm{min}}^i=T_{\mathrm{set}}^i$, the vehicle can travel through the entire approaching zone without deceleration. It will accelerate at a constant acceleration $a_+$ until it maintains $v_{\mathrm{max}}$ and then travels at uniform speed. Therefore, the acceleration of vehicle $i$ is calculated by:
\begin{equation*}
    a^i=\left\{
    \begin{aligned}
        a_+ \quad & \mbox{if }v^i<v_{\mathrm{max}},T_{\mathrm{min}}^i\geq T_{\mathrm{set}}^i ,\\
        a_- \quad & \mbox{if }v^i>v_{\mathrm{min}},T_{\mathrm{min}}^i < T_{\mathrm{set}}^i,\\
        0 \quad & \mbox{otherwise} .
    \end{aligned}
    \right
    .
\end{equation*}

We use a simple collision detection function to ensure the distance between vehicles is no less than the safety gap.
Vehicles that have traversed the crossing zone will travel uniformly at $v_{\mathrm{max}}$ and be deleted from $G$.

\subsubsection{Results and discussion}
We use average delay of vehicles in each simulation to evaluate the three sequencing policies. The delay experienced by a vehicle is the difference between the actual driving time and the theoretical driving time. The theoretical driving time is assuming there is no other vehicles and the vehicle travels uniformly with the maximum speed.

Fig. \ref{fig:SUMO_distance_time} shows the spatio-temporal trajectories of 10 vehicles under various sequencing policies. MS changes most vehicles' sequence of reaching the crossing zone, LQF changes few vehicles' order of reaching the crossing zone and FIFO does not change the vehicles' order of reaching the crossing zone. As expected, the MS policy leads to the minimal average delay (3.22 sec/veh), since it minimizes the number of switch-overs. Meanwhile, the LQF policy tends to alternate the direction to discharge and thus leads to the maximal average delay (4.79 sec/veh).

Fig. \ref{fig:SUMO_results} shows the average delay under various sequencing policies as well as under various demand patterns. In general, average delay increases with demand and rapidly blows up as the demand approaches a certain threshold. The thresholds are policy-dependent and largely consistent with the boundary of the theoretical capacity region given by Theorem~\ref{thm_two}. Specifically, for smaller demands, the impact of sequencing policy is less significant. As demand increases, the delay under LQF quickly rises and attains the congestion (i.e. red) domain. MS leads to the minimal delay and the maximal capacity, while FIFO gives intermediate performance. The simulation results are compliant with our theoretical analysis in Section \ref{sec_performance} that the intersection has the largest capacity under MS policy and the smallest capacity under LQF policy. 

%% file: sections/40_conclusion.tex
\section{Concluding remarks}
\label{sec_conclusion}

In this paper, we formulate the sequencing of vehicles at a signal-free intersection as a piecewise-deterministic Markov process. Our model captures the characteristics of typical sequencing policies and produces analytical guarantees on macroscopic performance metrics including capacity and delay. We use the Foster-Lyapunov stability theory to analyze the boundedness of traffic state and to obtain closed-form bounds for travel delay under various policies. In particular, we show that the min-switchover (resp. longer-queue-first) policy attains the best (resp. worst) system-wide performance. We also develop algorithms that implement various sequencing policies in practical settings and validate the theoretical results via micro-simulation-based experiments. This work provides useful tools and insights for intersection control. Possible future directions include extension to multi-origin-destination configurations and integration of learning-based methods to obtain adaptivity.

%%%%%%%%%%%%%%%%%%%%%%%%%%%%%%%
\section*{Acknowledgments}
The authors appreciate the discussion with Prof. Z.-P. Jiang and Prof. J. Chao at NYU.
The authors also appreciate the undergraduate students that helped with this work: H. Dai and Q. Dai contributed to the theoretical results, T. Yao, Y. Yao, J. Lin contributed to simulation.